\documentclass[pre,showpacs,showkeys,preprintnumbers,amsmath,amssymb,superscriptaddress,twocolumn]{revtex4}
\usepackage{graphicx}
\usepackage{amsfonts}
\usepackage{url}
\usepackage{epstopdf}
\usepackage{color}
\usepackage{tikz-cd}
\usepackage{comment}
\usepackage{bm}
\usepackage[colorlinks=true,linkcolor=blue,citecolor=red]{hyperref}%

\def\ve{\varepsilon}

\def\pa{\partial\Omega}
\def\P{{\mathbb P}}
\def\R{{\mathbb R}}

\def\x{\bm{x}}
\def\C{\mathcal{C}}
\def\diam{\mathrm{diam}}

\begin{document}

\title{First-passage times to anisotropic partially reactive targets}

\author{Adrien Chaigneau}
\email{adrien.chaigneau@protonmail.com}

\author{Denis~S.~Grebenkov}
 \email{denis.grebenkov@polytechnique.edu}
\affiliation{
Laboratoire de Physique de la Mati\`{e}re Condens\'{e}e (UMR 7643), \\ 
CNRS -- Ecole Polytechnique, IP Paris, 91120 Palaiseau, France}

\date{\today}

\begin{abstract}
We investigate restricted diffusion in a bounded domain towards a
small partially reactive target in three- and higher-dimensional
spaces.  We propose a simple explicit approximation for the principal
eigenvalue of the Laplace operator with mixed Robin-Neumann boundary
conditions.  This approximation involves the harmonic capacity and the
surface area of the target, the volume of the confining domain, the
diffusion coefficient and the reactivity.  The accuracy of the
approximation is checked by using a finite-elements method.  The
proposed approximation determines also the mean first-reaction time,
the long-time decay of the survival probability, and the overall
reaction rate on that target.  We identify the relevant length scale
of the target, which determines its trapping capacity, and investigate
its relation to the target shape.  In particular, we study the effect
of target anisotropy on the principal eigenvalue by computing the
harmonic capacity of prolate and oblate spheroids in various space
dimensions.  Some implications of these results in chemical physics
and biophysics are briefly discussed.
\end{abstract}





\maketitle

\section{Introduction}

Diffusion-controlled reactions play a central role in various
physical, chemical and biological phenomena
\cite{Rice,Lauffenburger,Redner,Schuss,Metzler,Oshanin,Grebenkov07,Benichou11,Bressloff13,Benichou14}.
At a single-molecule level, these processes are characterized by the
so-called first-passage time statistics.  In a typical setting, a
particle (e.g., a protein or an ion) diffuses inside a confining
domain and searches for a specific target (e.g., an enzyme or a
receptor) to react with.  The distribution of the reaction time (i.e.,
the first time instance at which the reaction occurs) depends on the
diffusive dynamics, the shapes of the domain and of the target, its
reactivity and location with respect to the starting position of the
diffusing particle
\cite{Collins49,Berg77,Sano79,Weiss86,Condamin07,Benichou08,Benichou10,Benichou10b,Rupprecht15,Godec16,Godec16b,Marshall16,Grebenkov16,Chechkin17,Lanoiselee18,Levernier19,Grebenkov20a}.
While this distribution can in general be obtained by solving the
Fokker-Planck equation with appropriate boundary conditions
\cite{Redner,Gardiner}, such a solution remains too formal and not
much informative, except for a few basic domains such as an interval,
concentric circles or spheres (see, e.g., \cite{Grebenkov18}).  

In the case of a {\it small} target, more explicit solutions are
available.  For instance, matched asymptotic methods can be employed
to compute the mean first-passage time, the smallest eigenvalue of the
governing Laplace operator and other characteristics of
diffusion-controlled reactions
\cite{Ozawa81,Mazya85,Ward93,Ward93b,Kolokolnikov05,Singer06a,Singer06b,Singer06c,Pillay10,Cheviakov10,Cheviakov11,Cheviakov12}
(see also review \cite{Holcman14} and references therein).  By a
different method based on pseudopotentials, Isaacson and Newby
developed a uniform asymptotic approximation of diffusion to a small
target \cite{Isaacson13}.  When the target is located on the boundary,
homogenization techniques can be applied
\cite{Zwanzig90,Grigoriev02,Berezhkovskii04,Berezhkovskii06,Muratov08,Dagdug16,Lindsay17,Bernoff18a,Bernoff18b}
(see also discussion in \cite{Grebenkov19d}).  In some geometric
settings, one can go further and develop self-consistent
approximations for the mean reaction time and its whole distribution
\cite{Grebenkov17a,Grebenkov17b,Grebenkov18a,Grebenkov19,Grebenkov21}.  
In the case of elongated domains, the original multi-dimensional
setting can be reduced to an effective one-dimensional problem that
admits explicit solutions \cite{Grebenkov20e,Grebenkov22a}.

When a small target is located inside a confining domain far from
reflecting boundaries, the shape of the target is generally ignored.
In fact, one often dealt with a spherical target, which is
characterized by a single length scale -- its diameter (or radius).
Even if a small sphere was replaced by a small cube or a small disk of
the same size, its reaction rate or trapping capacity for diffusing
particles would be modified insignificantly (see, e.g., examples in
\cite{Grebenkov22a}).  Several former studies were dedicated to the
impact of the target shape onto the trapping constant of
diffusion-limited reactions
\cite{Samson77,Cukier85,Berg85,Tsao02,McDonald04,Berezhkovskii07,Galanti16,Traytak18,Grimes18,Grebenkov18g,Piazza19}
and, more recently, onto the mean first-passage time
\cite{Grebenkov17b}.  Despite these works, the role of target
anisotropy in diffusion-controlled reactions remains poorly
understood.  In fact, if the target is elongated (e.g., cigar-shaped),
there are at least two relevant geometric length scales, namely, its
``length'' and ``width'', and identification of an appropriate
``size'' of the target is not clear.  In particular, if the ``length''
is fixed but the ``width'' vanishes, such a degenerated target (a
needle) becomes inaccessible to Brownian motion, i.e., its trapping
constant vanishes.  If the target is partially reactive
\cite{Collins49,Sano79,Zwanzig90,Berezhkovskii04,Galanti16,Lindsay17,Grebenkov17a,Bernoff18b,Grebenkov19d,Sano81,Shoup82,Sapoval94,Filoche99,Benichou00,Sapoval02,Grebenkov03,Grebenkov05,Grebenkov06a,Grebenkov06,Traytak07,Bressloff08,Singer08,Grebenkov10a,Grebenkov10b,Lawley15,Grebenkov15},
the anisotropy effect is even more sophisticated.

In this paper, we consider restricted diffusion in a bounded
$d$-dimensional domain towards a small partially reactive target.  We
focus on the the principal (smallest) eigenvalue $\lambda_1$ of the
Laplace operator, which is related to the reaction or trapping rate
and determines the mean first-reaction time and the decay rate of the
survival probability (see below).  We propose a simple approximation
for $\lambda_1$, which exhibits an explicit dependence on the target
reactivity.  This approximation allows us to identify the proper
trapping length of the target.  In order to analyze the effect of
target anisotropy, we will focus on spheroidal targets, for which the
trapping length can be computed exactly in any space dimension $d \geq
3$.  These targets are also used for numerical validation of the
proposed approximation.

The paper is organized as follows.  In Sec. \ref{sec:main}, we
formulate the general first-passage problem and derive an
approximation for the principal eigenvalue $\lambda_1$.  Section
\ref{sec:anisotropy} is devoted to the effect of target anisotropy
analyzed for spheroidal domains.  In Sec. \ref{sec:conclusion}, we
discuss the main results and their implications, as well as further
perspectives.  Appendices contain some technical derivations.

\section{Main results}
\label{sec:main}

\begin{figure}
\begin{center}
\includegraphics[width=45mm]{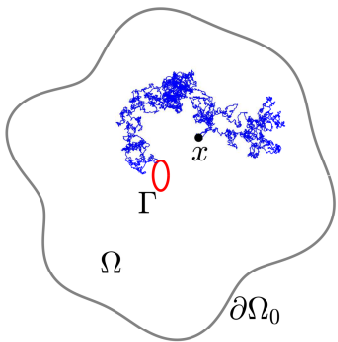} 
\end{center}
\caption{
A confining domain $\Omega$ with reflecting boundary $\pa_0$ (in
gray).  A particle diffuses (blue trajectory) from a starting point
$\x$ (black filled circle) towards an anisotropic target $\Gamma$ (in
red).  }
\label{fig:scheme}
\end{figure}

We consider a particle that starts from a point $\x$ and diffuses with
a diffusion coefficient $D$ inside a confining domain $\Omega \subset
\R^d$ with a smooth boundary $\pa = \pa_0 \cup \Gamma$ composed of two
disjoint parts: a reflecting ``outer'' boundary $\pa_0$ and a
partially reactive ``inner'' target $\Gamma$ with a reactivity
$\kappa$ (Fig. \ref{fig:scheme}).  Let $\tau$ denote the
first-reaction time, i.e., the instance when the particle reacts on
the target.  The survival probability of the particle (i.e., the
probability that the particle has not reacted up to time $t$),
$S_q(t|\x) = \P_{\x}\{ \tau > t\}$, satisfies the (backward) diffusion
equation
\begin{equation}
\partial_t S_q(t|\x) = D \Delta S_q(t|\x) \quad (\x\in\Omega),
\end{equation}
subject to the uniform initial condition $S_q(0|\x) = 1$ and mixed
Robin-Neumann boundary conditions \cite{Redner}:
\begin{equation}
\left\{  \begin{array}{rl}
\bigl(D\partial_n + \kappa \bigr) S_q(t|\x) = 0 & (\x\in \Gamma), \\
\partial_n S_q(t|\x) = 0 & (\x \in\pa_0).  \\  \end{array} \right.
\end{equation}
Here $\Delta$ is the Laplace operator, $\partial_n$ is the normal
derivative oriented outward the domain, and $q = \kappa/D$.  The
survival probability admits a general spectral decomposition
\cite{Redner,Gardiner}
\begin{equation}  \label{eq:Sq_spectral}
S_q(t|\x) = \sum\limits_{k=1}^\infty  e^{-Dt\lambda^{(q)}_k} u^{(q)}_k(\x) \int\limits_{\Omega} d\x' \, [u^{(q)}_k(\x')]^* ,
\end{equation}
where asterisk denotes the complex conjugate, and $\lambda^{(q)}_k$
and $u^{(q)}_k(\x)$ are the eigenvalues and orthonormal eigenfunctions
of the (negative) Laplace operator in $\Omega$, subject to mixed
Robin-Neumann boundary conditions:
\begin{subequations}
\begin{align}  \label{eq:un}
\Delta u^{(q)}_k(\x) + \lambda^{(q)}_k u^{(q)}_k(\x) & = 0  \quad (\x\in\Omega) ,\\  \label{eq:un_BC}
(\partial_n + q) u^{(q)}_k|_{\Gamma} & = 0 , \qquad \partial_n u^{(q)}_k|_{\pa_0} = 0. 
\end{align}
\end{subequations}
In general, the survival probability that fully characterizes the
distribution of the first-reaction time, exhibits a sophisticated
dependence on the shapes of the domain and of the target, on the
location of the starting point $\x$, on the diffusive dynamics (here,
the diffusivity $D$) and on the reaction mechanism (here, the
reactivity $\kappa$).  Various aspects of this dependence have been
investigated in the past
\cite{Benichou08,Isaacson13,Godec16b,Grebenkov18a,Grebenkov19,Grebenkov21,Lanoiselee18,Grebenkov18,Levernier19,Grebenkov20a,Grebenkov20d,Grebenkov20g,LeVot20,Grebenkov22}. 

In this paper, we focus on a common setting when the target is small
and located far away from the reflecting boundary $\pa_0$ of the
confining domain $\Omega$.  In this section, we will obtain the
following approximation to the principal (smallest) eigenvalue
$\lambda^{(q)}_1$ of the Laplace operator:
\begin{equation}  \label{eq:lambda1}
\lambda^{(q)}_1 \approx \frac{q|\Gamma|}{|\Omega|(1 + qL)}  \,,
\end{equation}
where
\begin{equation}  \label{eq:ell}
L = \frac{|\Gamma|}{C} 
\end{equation}
that we call the trapping length of the target.  Here $C$ is the
harmonic (or Newtonian) capacity of the target (see below), $|\Omega|$
is the Lebesgue measure of $\Omega$ (e.g., its volume in three
dimensions), and $|\Gamma|$ is the Lebesgue measure of the target
$\Gamma$ (e.g., its surface area in three dimensions).  In the
following, we describe the role of the trapping length $L$ and its
relation to the shape of the target.  We also check the accuracy of
this approximation and discuss immediate applications of this
approximation for the decay time, the mean first-reaction time, and
the reaction rate.

\subsection{Harmonic capacity}
\label{sec:derivation}

We start by recalling the notion of capacitance, which plays one of
the central roles in electrostatics.  The capacitance $C$ of an
isolated conductor $\C$ in $\R^3$ is the total charge on the
conductor's surface when it is maintained at unit potential
\cite{Jackson,Landau}.  In mathematical terms, the capacitance can be
defined as
\begin{equation}  \label{eq:C_def0}
C = \epsilon_0 \int\limits_{\R^3 \backslash \C} d\x \, |\nabla \Psi|^2 ,
\end{equation}
where $\epsilon_0 \approx 8.854\cdot 10^{-12}$~F/m is the vacuum
permittivity, and $\Psi(\x)$ is the (dimensionless) electric potential
outside the conductor satisfying
\begin{equation}  \label{eq:Psi_def}
\Delta \Psi(\x) = 0 \quad (\x\in\R^3 \backslash \C), \qquad 
\left\{ \begin{array}{l} \Psi|_{\partial \C} = 1 , \\ 
\lim\limits_{|\x|\to\infty} \Psi(\x) = 0. \\ \end{array} \right.
\end{equation}
For instance, the capacitance of a ball of radius $b$ is $4\pi
\epsilon_0 b$, which follows immediately from the classical radial
solution $\Psi(\x) = b/|\x|$.  In the following, we adopt a similar
notion of the harmonic (or Newtonian) capacity of a compact set $\C$
in $\R^d$
\cite{Landkof}:
\begin{equation}  \label{eq:C_def}
C = \int\limits_{\R^d \backslash \C} d\x \, |\nabla \Psi|^2 ,
\end{equation}
which is identical to Eq. (\ref{eq:C_def0}) but without the
fundamental constant $\epsilon_0$, and $\Psi(\x)$ satisfies the
Laplace equation in $\R^d\backslash \C$.  In particular, the capacity
of a ball of radius $b$ is $(d-2) \sigma_d b^{d-2}$, where
\begin{equation}  \label{eq:omegad}
\sigma_d = \frac{2 \pi^{d/2}}{\Gamma(d/2)}
\end{equation}
is the area of the $d$-dimensional unit ball, with $\Gamma(z)$ being
the Euler gamma function (not to be confused with our notation
$\Gamma$ for the target).  Note that some authors rescale the capacity
as $\hat{C} = \tfrac{1}{(d-2)\sigma_d} C$ to make the capacity of a
ball to be $b^{d-2}$.

According to Eq. (\ref{eq:Psi_def}), $\Psi(\x)$ can also be
interpreted as the probability of capture on the perfect target
$\Gamma = \partial\C$ of a Brownian particle started from $\x$.  The
perfect target refers to the Dirichlet boundary condition (i.e., $q =
\infty$) when the particle is captured by (or adsorbed on, or reacted on,
or killed on) the target $\Gamma$ upon their first encounter.  In
turn, $1 - \Psi(\x)$ is the steady-state survival (or escape)
probability of that particle (i.e., it is equal to the long-time limit
of $S_\infty(t|\x)$ in the case when there is no outer boundary
$\pa_0$).  Using the Green's formula, one can rewrite
Eq. (\ref{eq:C_def}) as
\begin{equation}  \label{eq:C_def1}
C = \int\limits_{\Gamma} d\x \, \partial_n \Psi .
\end{equation}
As a consequence, if there are many independent particles and their
concentration is maintained at $n_0$ at infinity, then $J_\infty = CD
n_0$ is the total steady-state diffusive flux onto the perfectly
absorbing target $\Gamma$, while $K_\infty = J_\infty/n_0 = CD$ is the
trapping constant of that target \cite{Berezhkovskii07}.  The analogy
between electrostatics and diffusion-controlled reactions have been
thoroughly employed in the past \cite{Redner}.  We emphasize that the
capacity, which is obtained by solving the Laplace equation in the
space outside the target, is the intrinsic property of that target.
In other words, there is no outer reflecting boundary here.

\subsection{Approximation for a perfect target}
\label{sec:perfect}

We explore yet another application of the capacity as a leading-term
approximation of the smallest eigenvalue $\lambda_1^{(\infty)}$ of the
Laplace operator in the presence of a perfect target ($q = \infty$)
for which the Robin boundary condition in Eq. (\ref{eq:un_BC}) is
reduced to the Dirichlet boundary condition
$(u^{(\infty)}_k)_{|\Gamma} = 0$.  This role of the capacity was
recognized already by Samarskii in 1948 \cite{Samarskii48}, but more
elaborate asymptotic analysis of the Dirichlet Laplace operator
eigenvalues was developed in \cite{Ozawa81,Mazya85,Cheviakov11}.  Here
the target $\Gamma$ is enclosed by an outer reflecting surface $\pa_0$
so that the confining domain $\Omega$ is bounded
(Fig. \ref{fig:scheme}).  We assume that the target is small as
compared to the confining domain $\Omega$, and is located far away
from the outer reflecting boundary $\pa_0$, i.e.,
\begin{equation}
\diam \{ \Gamma\} \ll  |\pa_0 - \Gamma| \leq \diam\{\Omega\} ,
\end{equation}
where $|\pa_0 - \Gamma|$ is the distance between sets $\pa_0$ and
$\Gamma$, and $\diam\{A\} = \sup_{\x_1,\x_2\in A} \bigl\{ |\x_1 -
\x_2|\bigr\}$ denotes the diameter of a set $A$.  Since
mathematical works \cite{Mazya85,Cheviakov11} were focused on the
three-dimensional setting (as well as the two-dimensional case in
\cite{Mazya85}), we briefly describe the general arguments valid for
any $d \geq 3$ (see the discussion for planar domains in
Sec. \ref{sec:conclusion}).

\begin{figure}
\begin{center}
\includegraphics[width=85mm]{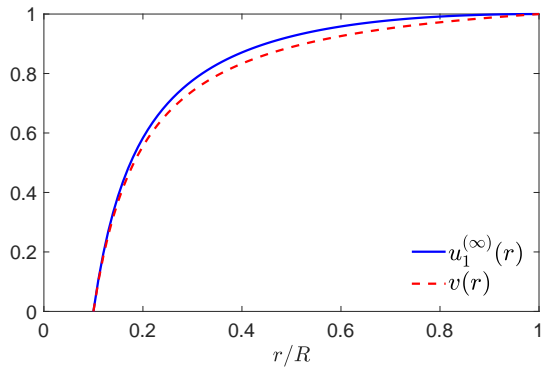} 
\end{center}
\caption{
The eigenfunction $u_1^{(\infty)}(r)$ of the Laplace operator in a
three-dimensional shell-like domain between two concentric spheres of
radii $b = 0.1$ and $R = 1$, with Dirichlet boundary condition on the
target, $u_1^{(\infty)}(b) = 0$, and Neumann boundary condition on the
outer sphere, $(\partial_r u_1^{(\infty)})(R) = 0$.  This
eigenfunction is known explicitly (see \cite{Grebenkov18} for details)
and depends only on the radial coordinate $r = |\x|$.  For comparison,
the harmonic function $v(r) = (1/b - 1/r)/(1/b - 1/R)$ satisfying
$v(b) = 0$ and $v(R) = 1$, is shown by dashed line.}
\label{fig:u1}
\end{figure}

Integrating Eq. (\ref{eq:un}) over $\x \in \Omega$ and using the
Green's formula, one gets
\begin{equation}  \label{eq:lambdak}
\lambda^{(\infty)}_1 = - \frac{\int\nolimits_{\Gamma} d\x \, \partial_n u^{(\infty)}_1(\x)}{\int\nolimits_{\Omega} d\x \, u^{(\infty)}_1(\x)} 
\end{equation}
(see, e.g., the review \cite{Grebenkov13} for other properties of
Laplacian eigenvalues and eigenfunctions).  As $\Gamma$ is small, the
numerator is small and thus the principal eigenvalue
$\lambda^{(\infty)}_1$ is close to $0$.  The associated eigenfunction
is therefore close to a constant function, $u^{(\infty)}_1(\x) \approx
u_0$, except for a boundary layer near the target; in particular, the
Neumann boundary condition at the outer reflecting boundary can be
replaced by the Dirichlet condition
$\bigl(u^{(\infty)}_1\bigr)_{|\pa_0} \approx u_0$.  In turn, the
eigenfunction $u^{(\infty)}_1(\x)$ vanishes on the target.  One can
thus approximate $u^{(\infty)}_1(\x)$ near the target by setting
$u^{(\infty)}_1(\x) \approx u_0 v(\x)$, where $v(\x)$ is the harmonic
function satisfying Dirichlet boundary conditions $v|_{\Gamma} = 0$
and $v|_{\pa_0} = 1$.  Substituting these approximations into
Eq. (\ref{eq:lambdak}), one gets
\begin{equation*}  
\lambda^{(\infty)}_1 \approx - \frac{\int\nolimits_{\Gamma} d\x \, \partial_n v(\x)}{\int\nolimits_{\Omega} d\x \, v(\x)} .
\end{equation*}
In the numerator, the integral is carried over the target $\Gamma$ so
that the function $v(\x)$ can be replaced by its limit $1- \Psi(\x)$,
which is obtained by moving the outer boundary $\pa_0$ to infinity.
In other words, a distant outer boundary $\pa_0$ does not much
influence the solution in the vicinity of the target.  In turn, the
denominator is the integral over the domain $\Omega$, in which $v(\x)$
is nearly constant, except for a vicinity of the small target.  We
replace therefore $v(\x)$ by $1$ here.  Upon these two approximations,
one gets
\begin{equation}  \label{eq:lambda1_inf}
\lambda^{(\infty)}_1 \approx \frac{\int\nolimits_{\Gamma} d\x \, \partial_n \Psi(\x)}{|\Omega|} = \frac{C}{|\Omega|} \,.
\end{equation}
Figure \ref{fig:u1} illustrates the behavior of the eigenfunction
$u_1^{(\infty)}(\x)$ and its approximation by $v(\x)$ for a shell-like
domain between two concentric spheres, for which these two functions
are known explicitly.

Moreover, Maz'ya {\it et al.} as well as Cheviakov and Ward provided
the next-order correction to this approximation in three dimensions
\cite{Mazya85,Cheviakov11}.  In our setting of a single target, their
result reads
\begin{equation}  \label{eq:lambda1_inf_corr}
\lambda_1^{(\infty)} \approx \frac{C'}{|\Omega|} \,,
\end{equation}
where $C'$ can be understood as a ``corrected'' capacity:
\begin{equation}  \label{eq:Cprime}
C' = C - C^2 R_N(\x_\Gamma,\x_\Gamma).
\end{equation}
Here $R_N(\x,\x')$ is the regular part of the Neumann Green's
function, and $\x_\Gamma$ is the location of (the center of) the
target $\Gamma$.  The Neumann Green's function is defined in the
confining domain {\it without any target} as
\begin{subequations}
\begin{align}
\Delta G_N(\x,\x') & = \frac{1}{|\Omega|} - \delta(\x-\x') \quad (\x\in\Omega), \\
\qquad \partial_n (G_N)|_{\pa_0} &= 0,  \qquad   \int\limits_{\Omega} d\x \, G_N(\x,\x') = 0,
\end{align}
\end{subequations}
and its regular part is
\begin{equation}
G_N(\x,\x') = \frac{1}{4\pi |\x-\x'|} + R_N(\x,\x').
\end{equation}
In other words, both $G_N(\x,\x')$ and $R_N(\x,\x')$ depend only on
the confining domain but are independent of the target.  For a
spherical domain of radius $R$, Cheviakov and Ward derived an explicit
expression for the Neumann Green's function and its regular part
\cite{Cheviakov11}.  In particular, they found
\begin{align}  \nonumber
R \, R_N(\x,\x) & = \frac{1}{4\pi (1-|\x|^2/R^2)} - \frac{1}{4\pi} \ln(1 - |\x|^2/R^2) \\
& + \frac{|\x|^2}{4\pi R^2} - \frac{7}{10\pi} \,.
\end{align}
For instance, if the target is located at the center, one has
$R_N({\bf 0},{\bf 0}) = - 9/(20\pi R)$.
We will discuss the accuracy of this approximation in
Sec. \ref{sec:anisotropy}.

\subsection{Global mean first-reaction time}

The next step consists in extending the above approximation to a
partially reactive target.  For this purpose, we employ the relation
between the smallest eigenvalue $\lambda_1^{(q)}$ and the so-called
global mean first-reaction time, $T_q$, which is defined as the volume
average of the mean first-reaction time $T_q(\x) = \langle
\tau\rangle$:
\begin{equation}
T_q = \frac{1}{|\Omega|} \int\limits_{\Omega} d\x \, T_q(\x) .
\end{equation}
In other words, the starting point is considered here as being
uniformly distributed inside the confining domain.  In turn, $T_q(\x)$
satisfies the boundary value problem \cite{Redner}
\begin{subequations}  \label{eq:Tq}
\begin{align}   \label{eq:Tq_bulk}
D \Delta T_q(\x) & = -1  \quad (\x\in\Omega), \\
(\partial_n + q)T_q(\x) & = 0 \quad (\x\in\Gamma), \\
\partial_n T_q(\x) & = 0 \quad (\x\in \pa_0).
\end{align}
\end{subequations}
The integral of Eq. (\ref{eq:Tq_bulk}) over $\x\in\Omega$ implies
\begin{align*}
-|\Omega| & = \int\limits_{\Omega} d\x \, D \Delta T_q(\x) = \int\limits_{\Gamma} d\x \, D (\partial_n T_q(\x)) \\
& = - \kappa \int\limits_{\Gamma} d\x \, T_q(\x)  ,
\end{align*}
i.e.,
\begin{equation}  \label{eq:Tq_Gamma}
\int\limits_{\Gamma} d\x \, T_q(\x) = \frac{|\Omega|}{\kappa} \,.
\end{equation}
Curiously, this integral does not depend on the diffusion coefficient
$D$.

To proceed, we multiply Eq. (\ref{eq:Tq_bulk}) by $T_\infty(\x)$,
subtract from it Eq. (\ref{eq:Tq_bulk}) with $q = \infty$ multiplied
by $T_q(\x)$, and integrate over $\x\in \Omega$:
\begin{align*}
& \bigl(T_q - T_\infty\bigr)|\Omega| = \int\limits_{\Omega} d\x \bigl(T_q(\x) - T_\infty(\x)\bigr) \\
& \quad = \int\limits_{\Omega} d\x \bigl(T_\infty(\x) \, D\Delta T_q(\x) 
- T_q(\x) \, D\Delta T_\infty(\x)\bigr) \\
& \quad = \int\limits_{\Gamma} d\x \bigl(\underbrace{T_\infty(\x)}_{=0} \, D\partial_n T_q(\x) 
- T_q(\x) \, D\partial_n T_\infty(\x)\bigr) .
\end{align*}
Note that $T_\infty(\x)$ can be obtained by integrating the
Dirichlet-Neumann Green's function, $G(\x|\x_0)$, satisfying
\begin{subequations}
\begin{align}
-D \Delta G(\x|\x_0) & = \delta(\x-\x_0) \quad (\x\in\Omega), \\
G(\x|\x_0) & = 0 \quad (\x\in\Gamma), \\
\partial_n G(\x|\x_0) & = 0 \quad (\x\in\pa_0), 
\end{align}
\end{subequations}
as follows:
\begin{equation}
T_\infty(\x) = \int\limits_\Omega d\x_0 \, G(\x|\x_0) .
\end{equation}
As a consequence, $-D \partial_n T_\infty(\x)$ turns out to be
proportional to the harmonic measure density \cite{Garnett},
$\omega(\x|\x_0)$, averaged over $\x_0$:
\begin{align}  \nonumber
\omega(\x) & \equiv \frac{1}{|\Omega|} \int\limits_{\Omega} d\x_0 \, \omega(\x|\x_0) 
= \frac{1}{|\Omega|} \int\limits_{\Omega} d\x_0 \, \bigl(-D \partial_n G(\x|\x_0)\bigr) \\
& = - \frac{1}{|\Omega|} \, D\partial_n T_\infty(\x).
\end{align}
We conclude that
\begin{equation}  \label{eq:Tq_global}
T_q = T_\infty + \int\limits_{\Gamma} d\x \, \omega(\x) \, T_q(\x) .
\end{equation}

This relation that we formally obtained from the boundary value
problem (\ref{eq:Tq}), has a clear probabilistic interpretation.  In
fact, the first-reaction time $\tau$ can be naturally split into two
contributions, $\tau = \tau_\infty + \tau_\Gamma$, where $\tau_\infty$
is the first-passage time to the target (i.e., the instance of the
first arrival onto the target), and $\tau_\Gamma$ is the
first-reaction time for a particle that was started on the target
$\Gamma$.  Accordingly, $T_\infty$ is the volume-averaged mean value
of $\tau_\infty$, whereas the second term in Eq. (\ref{eq:Tq_global})
is the target-surface-averaged mean value of $\tau_\Gamma$.  Indeed,
$\omega(\x)$ describes the probability density of the first arrival in
a vicinity of a boundary point $\x \in \Gamma$, from which the
particle continues to diffuse until the reaction on $\Gamma$.  In
other words, the second term is the average of $T_q(\x)$ over the
random first arrival point on $\Gamma$.  Qualitatively, the first and
the second terms represent respectively diffusion-limited and
reaction-limited contributions.  Expectedly, the first term depends on
the diffusion coefficient $D$ but is independent of the reactivity
$\kappa$.  In contrast, the second term formally depends on both $D$
and $\kappa$.  However, when the target is small, the volume-averaged
harmonic measure density $\omega(\x)$ is expected to be almost
uniform:
\begin{equation}
\omega(\x) \approx \frac{1}{|\Gamma|} \,.
\end{equation}
Substituting this approximation into Eq. (\ref{eq:Tq_global}) and
using Eq. (\ref{eq:Tq_Gamma}), we deduce
\begin{equation}  \label{eq:Tq_global2}
T_q \approx T_\infty + \frac{|\Omega|}{\kappa |\Gamma|} \,.
\end{equation}
In this approximation, the second term depends only on the reactivity
$\kappa$ but is independent of the diffusion coefficient $D$.  The
relation (\ref{eq:Tq_global2}) represents therefore two consecutive
additive contributions to the global mean first-reaction time: the
diffusion-limited contribution $T_\infty$ describing the transport of
the particle towards the target, and the reaction-limited contribution
due to the partial reactivity of the target.  These two complementary
contributions to the mean first-reaction time have been earlier
discussed for some symmetric domains \cite{Grebenkov17a,Grebenkov18}.
However, we are not aware of earlier derivations of this
representation in the general setting.  A similar separation of
diffusion-limited and reaction-limited contributions to the
steady-state diffusive flux $J_q$ can be already identified in the
Collins-Kimball solution for a spherical target of radius $b$ in
$\R^3$ \cite{Collins49} (see also \cite{Noyes61,Berg85}):
\begin{equation}  \label{eq:Jq_Collins}
\frac{4\pi b^2 n_0}{J_q} = \frac{b}{D} + \frac{1}{\kappa}  \,.
\end{equation}
In the same vein, two contributions to the impedance of a partially
blocking electrode have been identified and discussed
\cite{Sapoval94,Filoche99,Grebenkov03,Grebenkov06}.

\subsection{Partially reactive target}

To complete our derivation, we evaluate the global mean first-reaction
time $T_q$ according to its definition
\begin{equation}  \label{eq:Tq_def}
T_q = \int\limits_0^\infty dt \, t \, (-\partial_t S_q(t)) = \int\limits_0^\infty dt \, S_q(t),
\end{equation}
where $-\partial_t S_q(t)$ is the probability density of the
first-reaction time (averaged over the starting point), with
\begin{equation}
S_q(t) = \frac{1}{|\Omega|} \int\limits_{\Omega} d\x \, S_q(t|\x) = \sum\limits_{k=1}^\infty c^{(q)}_k \, e^{-Dt\lambda^{(q)}_k} \,,
\end{equation}
where
\begin{equation}
c^{(q)}_k = \frac{1}{|\Omega|} \left| \int\limits_{\Omega} d\x \, u^{(q)}_k(\x)\right|^2 ,
\end{equation}
and we used the spectral expansion (\ref{eq:Sq_spectral}).  Since
$S_q(0) = 1$, the positive coefficients $c^{(q)}_k$ can be understood
as the relative weights of the Laplacian eigenfunctions
$u_k^{(q)}(\x)$ in the survival probability $S_q(t)$.

When the target is small, the ground eigenfunction $u^{(q)}_1(\x)$ is
almost constant in $\Omega$ (except for a layer near the target, see
above).  As a consequence, other eigenfunctions, which are orthogonal
to $u^{(q)}_1$, have small contributions to $S_q(t)$, with $c^{(q)}_k
\approx 0$ for $k > 1$, whereas $c^{(q)}_1 \approx 1$ (see further
discussion in \cite{Grebenkov20h,Grebenkov22b}).  In other words,
\begin{equation}  \label{eq:Sq_approx}
S_q(t) \approx e^{-Dt\lambda^{(q)}_1} \,,
\end{equation}
that implies, according to Eq. (\ref{eq:Tq_def}), the following
approximation:
\begin{equation}  \label{eq:Tq_lambda1}
T_q \approx \frac{1}{D\lambda^{(q)}_1} \,.
\end{equation}
Substituting Eq. (\ref{eq:Tq_global2}) into this relation, we finally
arrive at
\begin{align*}
\lambda^{(q)}_1 & \approx \frac{1}{D (T_\infty + \frac{|\Omega|}{\kappa |\Gamma|})}
\approx \frac{1}{\frac{1}{\lambda^{(\infty)}_1} + \frac{|\Omega|}{q|\Gamma|}} 
\approx \frac{1}{\frac{|\Omega|}{C} + \frac{|\Omega|}{q|\Gamma|}} 
\end{align*}
that implies the announced expression (\ref{eq:lambda1}).  This
relation can also be expressed in terms of the global mean
first-reaction time from Eq. (\ref{eq:Tq_lambda1}):
\begin{equation}
T_q \approx \frac{|\Omega|}{|\Gamma|} \, \biggl(\frac{L}{D} + \frac{1}{\kappa}\biggr) \,,
\end{equation}
which represent the sum of diffusion-limited and reaction-limited
contributions.  Accordingly, $1/T_q$ can be interpreted as the overall
reaction rate, while $T_q$ is also the decay time of the survival
probability at long times, $S_q(t|\x) \propto e^{-Dt\lambda_1^{(q)}}$,
see Eq. (\ref{eq:Sq_spectral}).  Note that this asymptotic relation
was employed to compute the principal eigenvalue numerically via
estimating the survival probability \cite{Lejay07}.

Moreover, the principal eigenvalue $\lambda_1^{(q)}$ can be used to
determine the steady-state diffusive flux and the trapping constant of
a small target.  In fact, the probability density $H_q(t|\x)$ can also
be understood as the probability flux onto the target from a fixed
point $\x$.  At long times, the spectral expansion
(\ref{eq:Sq_spectral}) implies
\begin{equation}
H_q(t|\x) \approx D \lambda_1^{(q)} \, e^{-D\lambda_1^{(q)} t} \, u_1^{(q)}(\x) \int\limits_{\Omega} d\x' \, u_1^{(q)}(\x') .
\end{equation}
As in Sec. \ref{sec:perfect}, one can argue that $u_1^{(q)}(\x)$ is
nearly constant for any $\x$ far from the target so that
\begin{equation}  \label{eq:Hq_asympt}
H_q(t|\x) \approx D \lambda_1^{(q)} \, e^{-D\lambda_1^{(q)} t}  ,
\end{equation}
where we used the $L_2(\Omega)$-normalization of $u_1^{(q)}(\x)$.  If
there are many independent particles with a concentration $n_0$, their
total diffusive flux onto the target is $J_q(t) \approx n_0 |\Omega|
H_q(t|\x)$.  Expectedly, this flux vanishes in the long time limit
because all particles that were initially present in a bounded domain,
react on the target.  However, if the target is very small, there is
an intermediate range of times, for which Eq. (\ref{eq:Hq_asympt})
holds but $D\lambda_1^{(q)} t \ll 1$, so that
\begin{equation}  \label{eq:Jq_inf}
J_q \approx n_0 |\Omega| D \lambda_1^{(q)} \approx n_0 D \frac{q|\Gamma|}{1 + q L} \,,
\end{equation}
where we used our approximation (\ref{eq:lambda1}) for
$\lambda_1^{(q)}$.  This is an extension of the Collins-Kimball's
relation (\ref{eq:Jq_Collins}) that was derived for a spherical
target.  While we derived the approximate relation (\ref{eq:Jq_inf})
by considering the limit of very small targets, one could
alternatively fix the target size and move the outer boundary $\pa_0$
to infinity.  In other words, this relation is applicable to a bounded
target of any size in $\R^d$ (i.e., without $\pa_0$).  Dividing the
total flux by $n_0$ yields the trapping constant:
\begin{equation}
K_q \approx D \frac{q|\Gamma|}{1 + q L} \,.
\end{equation}
In the limit $q\to \infty$, we retrieve the known approximations
$J_\infty \approx n_0 D C$ and $K_\infty \approx CD$ for perfectly
reactive targets that we mentioned in Sec. \ref{sec:derivation}.

In summary, the approximate relation (\ref{eq:lambda1}) relies on
three approximations (\ref{eq:lambda1_inf}, \ref{eq:Tq_global2},
\ref{eq:Sq_approx}), which are all based on the assumption of the
target smallness.  We stress that the above derivation does not
pretend to mathematical rigor.  A more rigorous derivation of
Eq. (\ref{eq:lambda1}) presents an interesting perspective.

\section{Target anisotropy}
\label{sec:anisotropy}

In former works on partially reactive targets
\cite{Collins49,Sano79,Zwanzig90,Berezhkovskii04,Galanti16,Lindsay17,Grebenkov17a,Bernoff18b,Grebenkov19d,Sano81,Shoup82,Sapoval94,Filoche99,Benichou00,Sapoval02,Grebenkov03,Grebenkov05,Grebenkov06a,Grebenkov06,Traytak07,Bressloff08,Singer08,Grebenkov10a,Grebenkov10b,Lawley15,Grebenkov15},
the reaction length $1/q = D/\kappa$ was generally compared to a
``typical size'' of the target, without providing its definition.  For
a spherical (or, more generally, ``roundish'') target, there is a
single geometric length scale, its diameter (or radius), which is
naturally compared with $1/q$.  In turn, when the target has an
approximately isotropic shape but a rough boundary, other geometric
length scales can emerge.  For instance, in the study of steady-state
diffusion of oxygen molecules towards the acinar surface in the lungs,
Sapoval {\it et al.} introduced the relevant length scale $L_S =
|\Gamma|/\diam\{\Gamma\}$ as the surface area of the target divided by
its diameter \cite{Sapoval02}.  As the surface area of a compact
target with a rough (e.g., fractal-like) boundary can be extremely
large, the length $L_S$ can be orders of magnitude larger than the
diameter itself.

The explicit approximation (\ref{eq:lambda1}) allows us to identify
the relevant length scale of a small target in a more general setting
and beyond the steady-state regime.  The trapping length $L =
|\Gamma|/C$ generalizes the above length $L_S$ to anisotropic targets
and in higher dimensions.  These two lengths are comparable for a
nearly isotropic target in three dimensions because the capacity of
such a target is comparable to its diameter.  In this section, we
investigate how the target anisotropy affects the trapping length $L$
and therefore various properties of diffusion-reaction processes.

\subsection{Prolate spheroids}

\begin{figure}[ht!]
\begin{center}
\includegraphics[width=85mm]{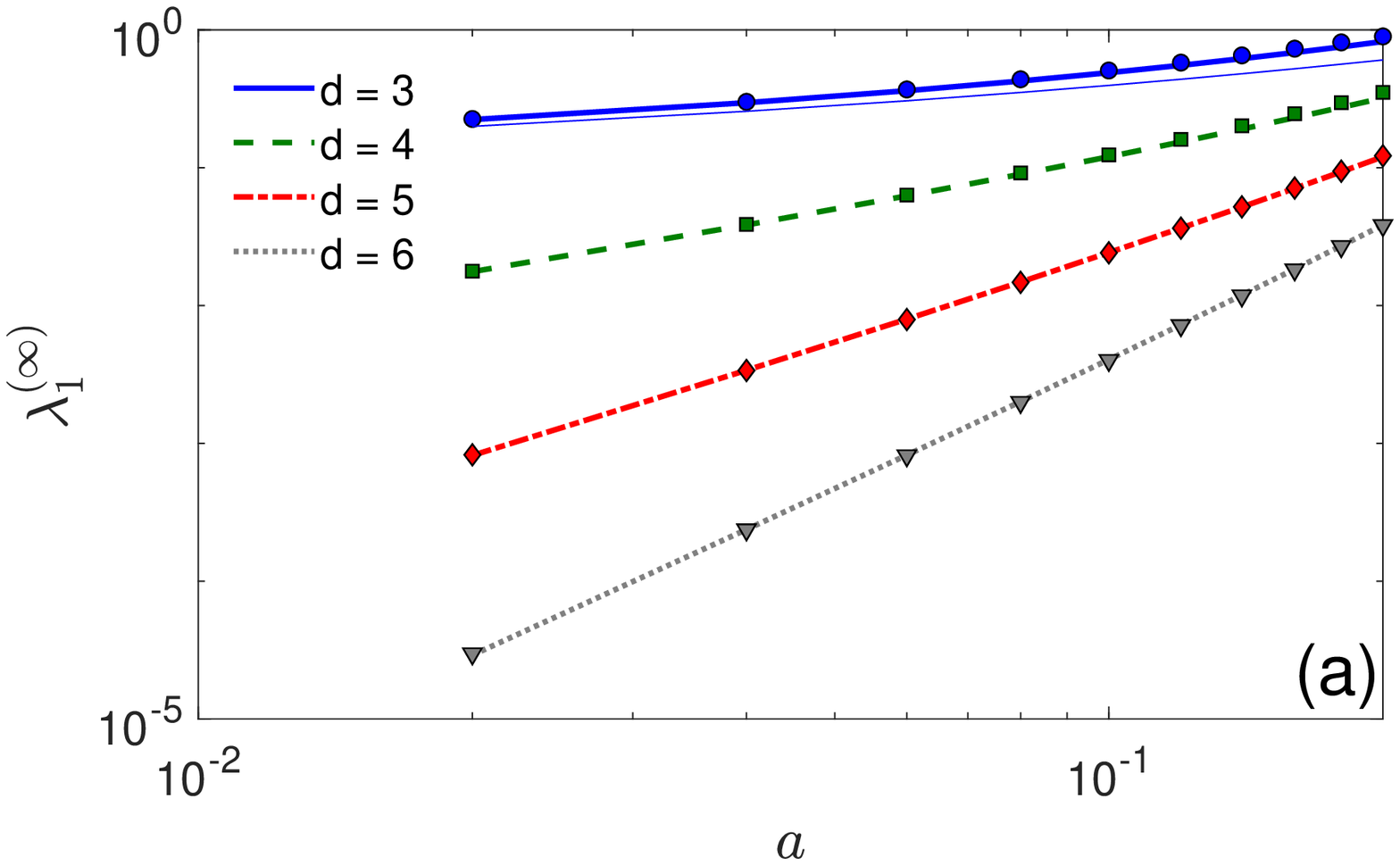} 
\includegraphics[width=85mm]{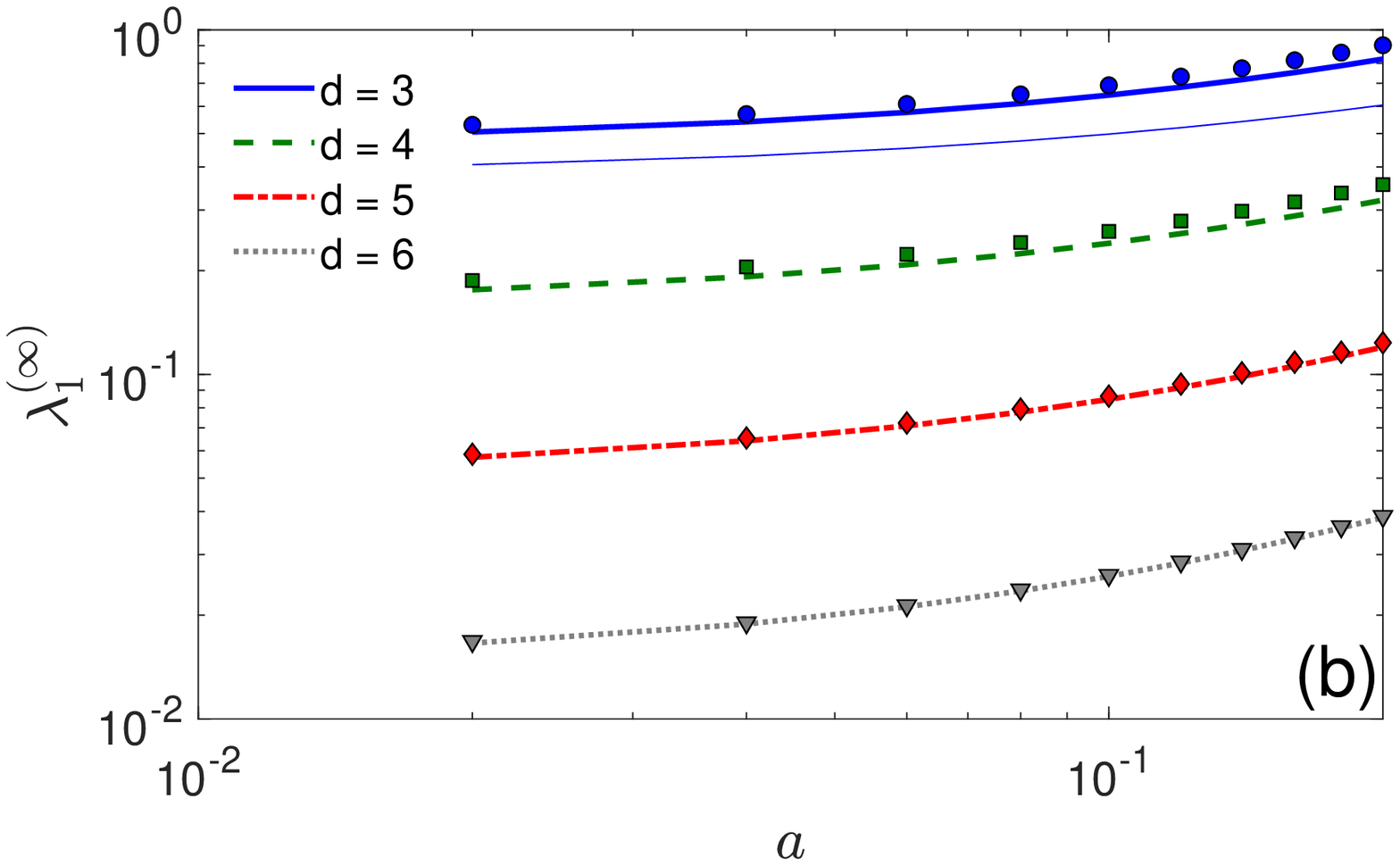} 
\includegraphics[width=85mm]{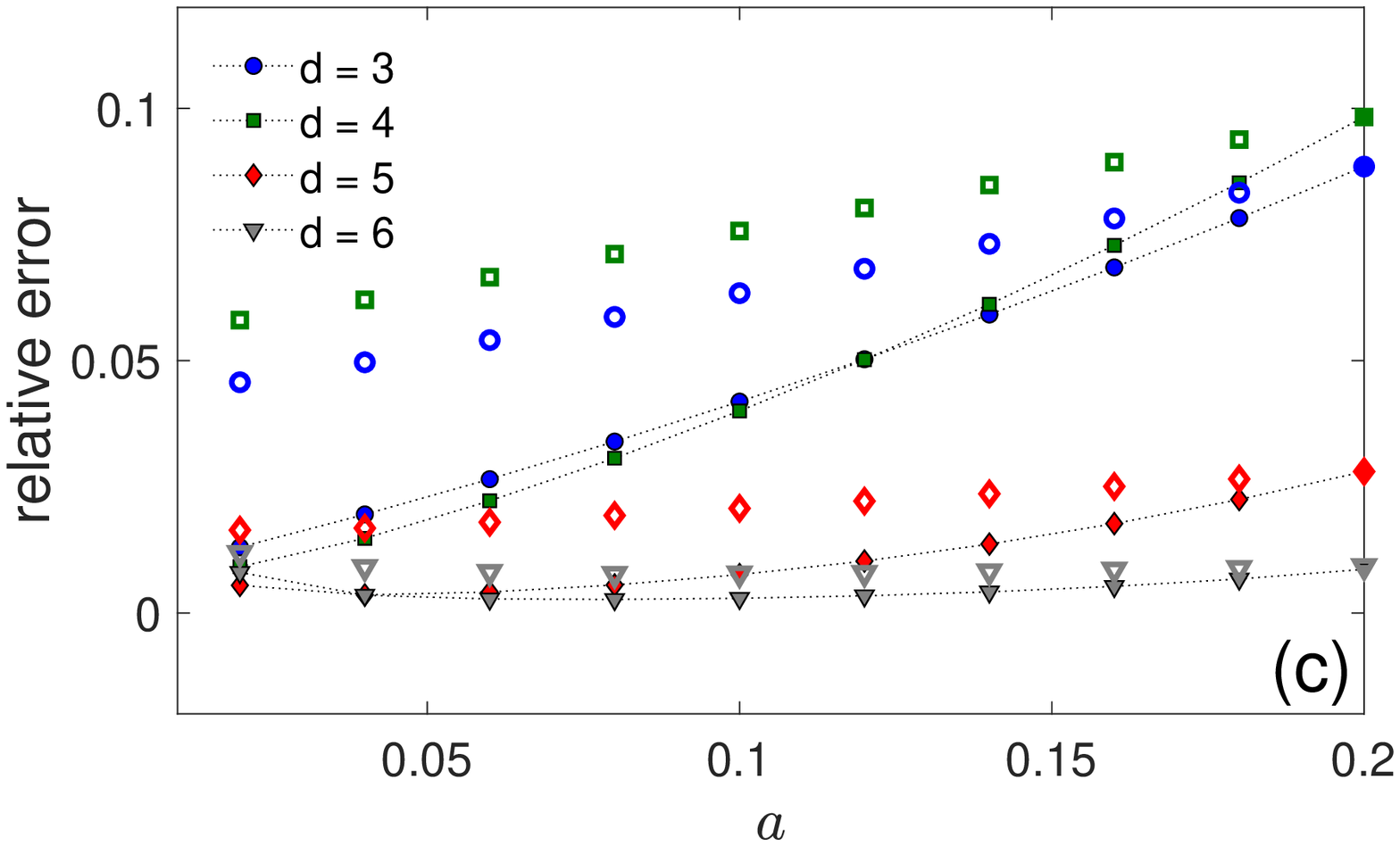} 
\end{center}
\caption{
{\bf (a,b)} The principal eigenvalue $\lambda_1^{(\infty)}$ of the
Laplace operator for a perfectly reactive prolate {\bf (a)} and oblate
{\bf (b)} spheroidal target with semi-axes $a \leq b = 0.2$ surrounded
by a concentric reflecting spherical surface of radius $R = 1$.
Symbols present the numerical computation by a finite-elements method
(see Appendix \ref{sec:FEM}), whereas thick lines show the approximate
relation (\ref{eq:lambda1_inf}).  In three dimensions, thick blue line
presents the improved approximation (\ref{eq:lambda1_inf_corr}) with
the ``corrected'' capacity $C'$ from Eq. (\ref{eq:Cprime}), whereas
thin blue line indicates the leading-order approximation
(\ref{eq:lambda1_inf}).  {\bf (c)} The relative error of the above
approximations shown by filled symbols for prolate spheroids and by
empty symbols for oblate spheroids in $\R^d$ with $d = 3,4,5,6$ (see
the legend).  Note that a minor increase of the relative error for $d
= 6$ at small $a$ can be a numerical artifact due to an insufficient
mesh size.}
\label{fig:lambda1}
%
\end{figure}

We model an elongated target by the surface of a $d$-dimensional
prolate spheroid (i.e., an ellipsoid of revolution) with the single
major semi-axis $b$ along the $d$-th coordinate, and equal minor
semi-axes $a < b$:
\begin{equation}  \label{eq:Gamma_prolate}
\Gamma_{a,b} = \left\{ (x_1,\ldots,x_d) \in \R^d : \frac{x_1^2}{a^2} + \ldots + \frac{x_{d-1}^2}{a^2} + \frac{x_d^2}{b^2} = 1\right\}.
\end{equation}
The capacity of a prolate spheroid in three dimensions is well known
\cite{Landau}:
\begin{equation}  \label{eq:C3}
C^{(3)}_{a,b} = \frac{8\pi c}{\ln \bigl(\frac{1 + c/b}{1 - c/b}\bigr)} \,,
\end{equation}
where $c = \sqrt{b^2 - a^2}$.  In the limit $a\to b$, this relation is
reduced to the classical capacity of a ball of radius $b$:
$C^{(3)}_{b,b} = 4\pi b$.  An extension of this result to higher
dimensions was discussed in \cite{Tee05}.  In Appendix
\ref{sec:prolate}, we describe this extension and obtain the
following compact expression:
\begin{equation}  \label{eq:Cd_prolate}
C^{(d)}_{a,b} = \frac{(d-2) \sigma_d \, b \, a^{d-3}}{~ _2F_1\bigl(\frac{1}{2}, 1 ; \frac{d}{2}; 1 - \frac{a^2}{b^2} \bigr)} \,,
\end{equation}
where $_2F_1(a,b;c;z)$ is the hypergeometric function, and $\sigma_d$
is given by Eq. (\ref{eq:omegad}).  For even dimensions, one gets
particularly simple expressions, e.g.
\begin{subequations}
\begin{align}
C^{(4)}_{a,b} & = 2\pi^2 a(a+b) , \\
C^{(6)}_{a,b} & = \frac{3\pi^3 a^3 (a+b)^2}{2a+b} \,.
\end{align}
\end{subequations}
In the limit $a\to b$, one retrieves the capacity of the ball:
$C^{(d)}_{b,b} = (d-2)\sigma_d b^{d-2}$.  In turn, in the opposite
limit of highly anisotropic targets, $a\to 0$, one can use the Euler's
identity  to get in the leading order:
\begin{equation}  \label{eq:Cd_a0}
C^{(d)}_{a,b} \approx (d-3) \sigma_d \, b \, a^{d-3}  \qquad (d > 3).
\end{equation}  
For $d = 3$, Eq. (\ref{eq:C3}) yields 
\begin{equation}
C^{(3)}_{a,b} \approx \frac{4\pi b}{\ln(b/a)} \,, 
\end{equation}
i.e., the capacity vanishes very slowly.  When the target is
surrounded by a concentric spherical surface $\pa_0$ of radius $R$,
the volume of the confining domain is
\begin{equation}
|\Omega| =  \frac{\pi^{d/2}}{\Gamma(d/2+1)} \bigl(R^d - ba^{d-1}\bigr).
\end{equation}

Figure \ref{fig:lambda1}(a) illustrates the behavior of the principal
eigenvalue $\lambda_1^{(\infty)}$ for a perfectly reactive target ($q
= \infty$).  On this log-log plot, one sees the expected power-law
dependence on the minor semi-axis $a$.  Our approximation
(\ref{eq:lambda1_inf}) is least accurate in three dimensions (thin
blue curve) and gets more and more accurate as the space dimension $d$
increases.  Note that the use of the ``corrected'' capacity $C'$ in
Eq. (\ref{eq:Cprime}) instead of $C$ significantly improves the
accuracy of the approximation in three dimensions (thick blue curve).
In Fig. \ref{fig:lambda1}(c), filled symbols show the relative error
of the approximation (\ref{eq:lambda1_inf}) for $d > 3$ and of
Eq. (\ref{eq:lambda1_inf_corr}) for $d = 3$.  For the considered major
semi-axis $b = 0.2$, the relative error does not exceed $10\%$.

The surface area of prolate spheroids is also discussed in Appendix
\ref{sec:prolate}:
\begin{equation}  \label{eq:area_prolate}
|\Gamma_{a,b}^{(d)}| = \sigma_d \, a^{d-2} \, b ~ _2F_1\biggl(\frac12,-\frac{1}{2}; \frac{d}{2}; 1 - \frac{a^2}{b^2}\biggr).
\end{equation}
As $a\to 0$, one gets in the lowest order
\begin{equation} 
|\Gamma_{a,b}^{(d)}| \approx 2\pi^{d/2} a^{d-2} b \, \frac{\Gamma(d/2)}{\Gamma(\frac{d-1}{2}) \Gamma(\frac{d+1}{2})} \,.
\end{equation}
Substituting Eqs. (\ref{eq:Cd_prolate}, \ref{eq:area_prolate}) into
Eq. (\ref{eq:ell}), we get the trapping length
\begin{equation}  \label{eq:elld_prolate}
L = \frac{a}{d-2} \, _2F_1\biggl(\frac12,-\frac{1}{2}; \frac{d}{2}; 1 - \frac{a^2}{b^2}\biggr)
~ _2F_1\biggl(\frac{1}{2}, 1 ; \frac{d}{2}; 1- \frac{a^2}{b^2} \biggr).
\end{equation}
For a spherical target ($a = b$), one retrieves $L = b/(d-2)$.  In
the opposite limit $a\to 0$ of highly anisotropic targets, we obtain
\begin{subequations}
\begin{align}
L & \approx a \, \frac{\Gamma^2\bigl(\frac{d}{2}\bigr)}{(d-3)\Gamma\bigl(\frac{d-1}{2}\bigr) \, \Gamma\bigl(\frac{d+1}{2}\bigr)} 
\qquad (d > 3), \\
L &\approx \frac{\pi}{4} \, a \ln(b/a)   \hskip 27.5mm (d = 3).
\end{align}
\end{subequations}
In both cases, the length scale $L$ vanishes, and the trapping
capacity of a very thin target becomes essentially reaction-limited
for any finite reactivity: $\lambda_1^{(q)} \approx q
|\Gamma|/|\Omega|$.

\begin{figure}
\begin{center}
\includegraphics[width=85mm]{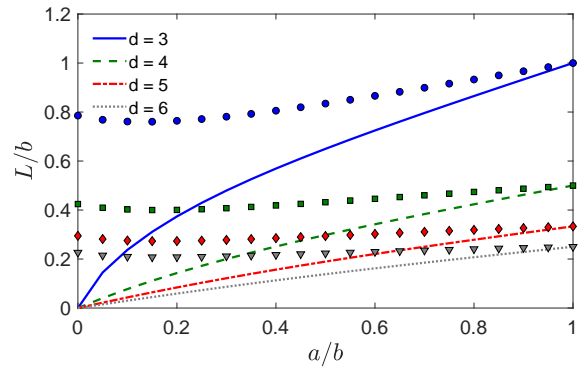} 
\end{center}
\caption{
The trapping length $L$ from Eqs. (\ref{eq:elld_prolate},
\ref{eq:elld_oblate}) of prolate (lines) and oblate (symbols)
spheroids for several dimensions $d$.  Note that $L/b = 1/(d-2)$ at
$a/b = 1$. }
\label{fig:ell}
\end{figure}

The dependence (\ref{eq:elld_prolate}) of the trapping length $L$ on
the aspect ratio $a/b$ is shown by lines in Fig. \ref{fig:ell}.  A
linear scaling of $L$ with $a$ is observed in all dimensions $d > 3$,
whereas the curve for $d = 3$ exhibits a linear scaling with a
logarithmic correction.

\begin{figure}
\begin{center}
\includegraphics[width=85mm]{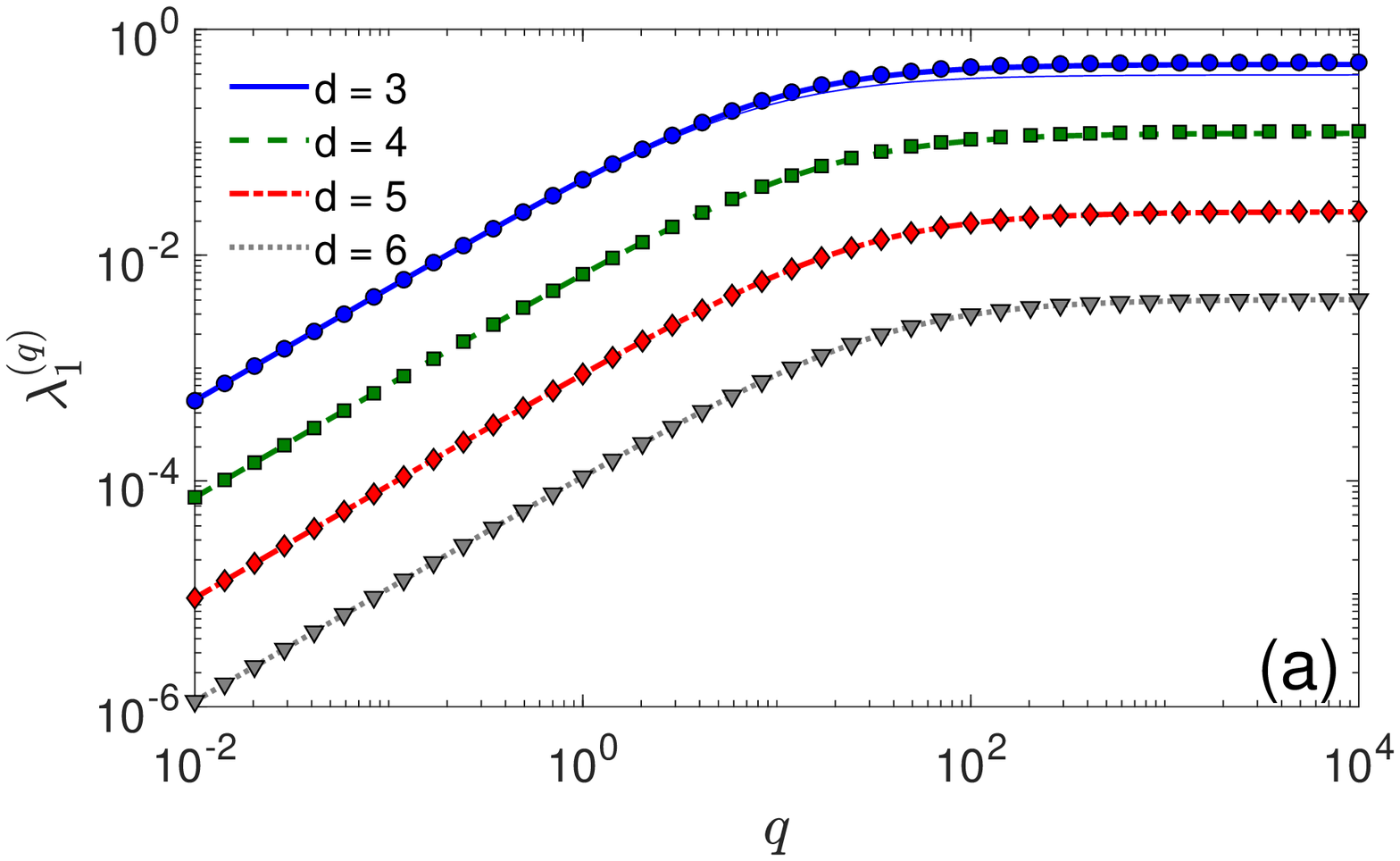} 
\includegraphics[width=85mm]{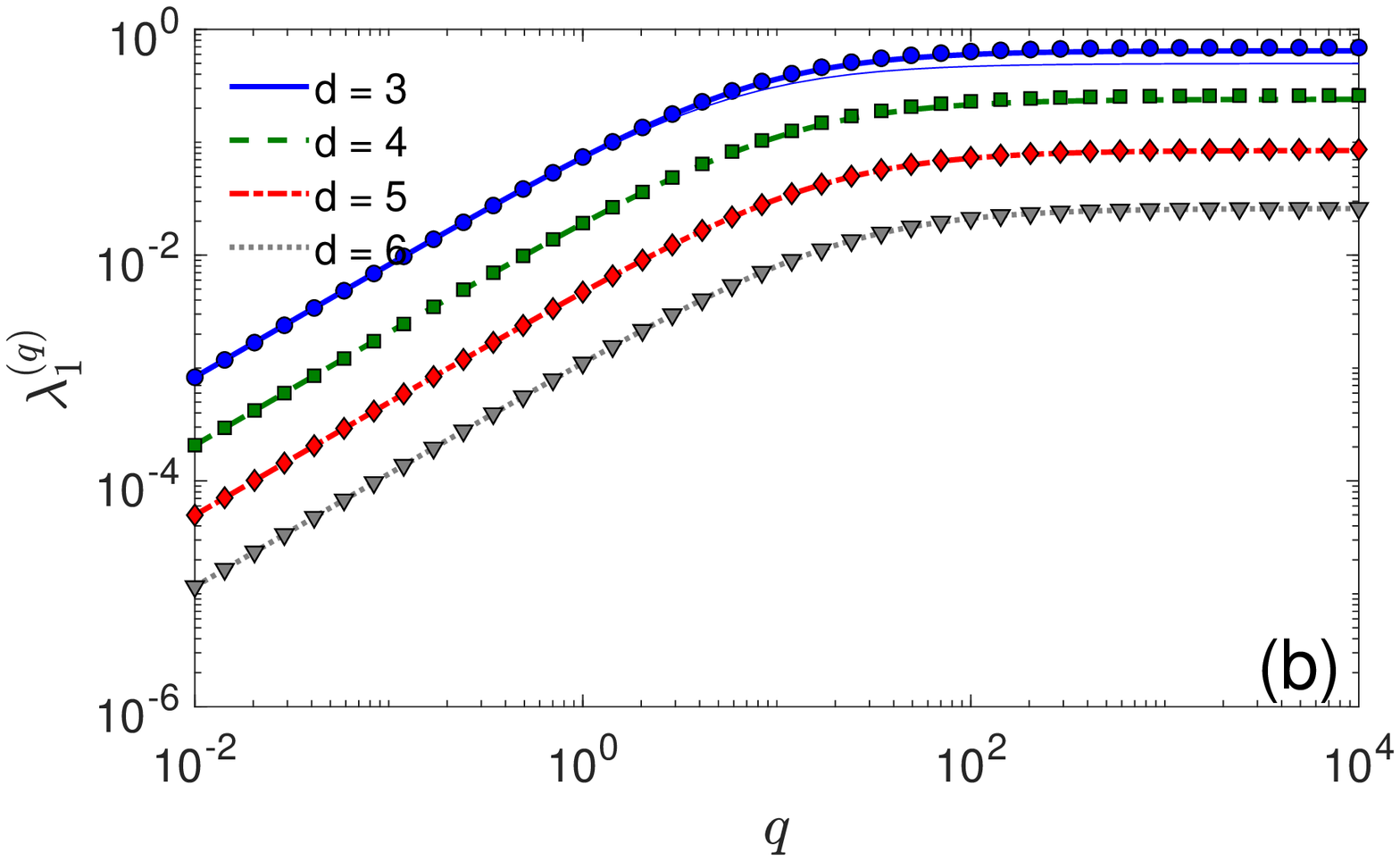} 
\end{center}
\caption{
The principal eigenvalue $\lambda_1^{(q)}$ as a function $q$ for
prolate {\bf (a)} and oblate {\bf (b)} spheroidal targets with
semi-axes $a = 0.1$ and $b = 0.2$ surrounded by a reflecting
concentric spherical surface of radius $R = 1$.  Symbols present the
numerical computation by a finite-elements method (see Appendix
\ref{sec:FEM}), whereas thick lines show the approximate relation
(\ref{eq:lambda1}).  In three dimensions, thick blue line presents
Eq. (\ref{eq:lambda1}) with the ``corrected'' capacity $C'$ from
Eq. (\ref{eq:Cprime}), whereas thin blue line corresponds to the
capacity $C$.  The trapping length $L$ given by
Eqs. (\ref{eq:elld_prolate},
\ref{eq:elld_oblate}) is $0.1300$, $0.0596$, $0.0379$, $0.0276$ for
prolate spheroids, and $0.1669$, $0.0864$, $0.0588$, $0.0447$ for
oblate spheroids, with $d = 3,4,5,6$, respectively.  }
\label{fig:lambda_q}
\end{figure}

Figure \ref{fig:lambda_q}(a) shows the principal eigenvalue
$\lambda_1^{(q)}$ as a function of $q$ for a prolate spheroid of a
fixed aspect ratio $a/b = 0.5$.  One sees that our approximation
(\ref{eq:lambda1}) is very accurate over a broad range of $q$ values
and all dimensions $d \geq 3$.

\subsection{Oblate spheroids}

A flattened target is modeled by the surface of a $d$-dimensional
oblate spheroid with the single minor semi-axis $a$ along the $d$-th
coordinate, and equal major semi-axes $b > a$:
\begin{equation}  \label{eq:Gamma_oblate}
\tilde{\Gamma}_{a,b} = \left\{ (x_1,\ldots,x_d) \in \R^d : \frac{x_1^2}{b^2} + \ldots + \frac{x_{d-1}^2}{b^2} + \frac{x_d^2}{a^2} = 1\right\} .
\end{equation}
The capacity of an oblate spheroid in three dimensions
is well known \cite{Landau}:
\begin{equation}
\tilde{C}^{(3)}_{a,b} = \frac{4\pi c}{\cos^{-1}(a/b)} \,.
\end{equation}
In the limit $a\to b$, one retrieves the capacity of the ball of
radius $b$; in the opposite limit $a\to 0$, this relation yields the
well-known result for the capacity of the disk of radius $b$:
$\tilde{C}^{(3)}_{0,b} = 8\pi b$.

In Appendix \ref{sec:oblate}, we recall the derivation of the capacity
in higher dimensions and derive the following compact expression
\begin{equation}  \label{eq:Coblate}
\tilde{C}^{(d)}_{a,b} = \frac{(d-2) \sigma_d \, b^{d-2}}{_2F_1\bigl(\frac{1}{2}, \frac{d-2}{2}; \frac{d}{2}; 1 -a^2/b^2\bigr)} \,.
\end{equation}
For even dimensions, one gets particularly simple relations, e.g.,
\begin{subequations}
\begin{align}
\tilde{C}^{(4)}_{a,b} & = 2\pi^2 b(a+b), \\
\tilde{C}^{(6)}_{a,b} & = \frac{3\pi^3 b^3 (a+b)^2}{2b+a} \,.
\end{align}
\end{subequations}
As $a\to 0$, the capacity reaches a finite limit:
\begin{equation}  \label{eq:Coblate0}
\tilde{C}^{(d)}_{0,b} = \frac{(d-2) \sigma_d \, b^{d-2} \Gamma(\frac{d-1}{2})}{\Gamma(\frac{d}{2}) \sqrt{\pi}} \,.
\end{equation}
In contrast to the case of infinitely thin elongated targets
(cf. Eq. (\ref{eq:Cd_a0})), flattened targets remain accessible to
Brownian motion.  When the target is surrounded by a concentric
spherical surface $\pa_0$ of radius $R$, the volume of the confining
domain is
\begin{equation}
|\tilde{\Omega}| =  \frac{\pi^{d/2}}{\Gamma(d/2+1)} \bigl(R^d - ab^{d-1}\bigr).
\end{equation}
The accuracy of the approximation (\ref{eq:lambda1}) for perfectly
reactive oblate targets is illustrated in Fig. \ref{fig:lambda1}(b).
As for elongated targets, the approximation is least accurate for $d =
3$ and gets more and more accurate as $d$ increases.  Its relative
error is shown in Fig. \ref{fig:lambda1}(c) by empty symbols.

The surface area of oblate spheroids is discussed in Appendix
\ref{sec:oblate}:
\begin{equation}  \label{eq:area_oblate}
|\tilde{\Gamma}_{a,b}^{(d)}| = \sigma_d \, b^{d-1} ~ _2F_1\biggl(\frac{d-1}{2},-\frac{1}{2}; \frac{d}{2}; 1 - \frac{a^2}{b^2}\biggr).
\end{equation}
In the limit $a\to 0$, one gets
\begin{equation}  \label{eq:Gamma_oblate_a0}
|\tilde{\Gamma}_{0,b}^{(d)}| =  b^{d-1} \, \frac{2 \pi^{(d-1)/2}}{\Gamma(\frac{d+1}{2})} \,.
\end{equation}
For instance, one retrieves the surface area of two-sided disk for $d
= 3$: $|\tilde{\Gamma}_{0,b}^{(3)}| = 2\pi b^2$ (it is twice bigger
than the area of the disk because there are two faces).

Substituting Eqs. (\ref{eq:Coblate}, \ref{eq:area_oblate}) into
Eq. (\ref{eq:ell}), we get the trapping length:
\begin{align}  \nonumber
L & = \frac{b}{d-2} ~ _2F_1\biggl(\frac{d-1}{2},-\frac{1}{2}; \frac{d}{2}; 1 - \frac{a^2}{b^2}\biggr) \\     \label{eq:elld_oblate}
& \times  ~ _2F_1\biggl(\frac{1}{2}, \frac{d-2}{2}; \frac{d}{2}; 1 - \frac{a^2}{b^2}\biggr)  \,.
\end{align}
In contrast to the case of prolate spheroids, the trapping length here
remains of the order of $b$ for any $a$, ranging from
\begin{equation}
L = \frac{b}{d-2}  \, \frac{\Gamma^2\bigl(\frac{d}{2}\bigr)}{\Gamma\bigl(\frac{d-1}{2}\bigr) \, \Gamma\bigl(\frac{d+1}{2}\bigr)} 
\qquad (a=0)
\end{equation}
to $L = b/(d-2)$ at $a = b$.  This behavior is shown in
Fig. \ref{fig:ell} by symbols.  Curiously, the dependence is not
monotonous but variations of $L$ with $a/b$ are insignificant,
particularly at larger $d$.  We conclude that flattening the target
does not almost change its trapping capacity.  The accuracy of the
approximation (\ref{eq:lambda1}) for a partially reactive oblate
target is illustrated in Fig. \ref{fig:lambda_q}(b).

\section{Discussion and conclusion}
\label{sec:conclusion}

In this paper, we investigated restricted diffusion inside a bounded
domain towards a partially reactive target.  Our first result is a
simple explicit approximation (\ref{eq:lambda1}) for the principal
eigenvalue of the Laplace operator with mixed Robin-Neumann boundary
conditions.  This approximation involves very basic geometric
characteristics such as the volume of the confining domain $|\Omega|$,
the surface area of the target $|\Gamma|$, and its harmonic capacity
$C$.  The dependence on the physical transport parameters, the
diffusion coefficient $D$ and the reactivity $\kappa$, is fully
explicit.  Even though the derivation of Eq. (\ref{eq:lambda1})
involved three approximations, all of them were based on the smallness
of the target and its distant location from the reflecting boundary.
A comparison with a numerical solution by a finite-elements method
showed that the approximation is getting more and more accurate as the
space dimension increases.  In three dimensions, the use of the
``corrected'' capacity $C'$ allows one to get accurate results as
well.  As the principal eigenvalue $\lambda_1^{(q)}$ determines
several characteristics of diffusion-controlled reactions, the
proposed approximation opens access to them in a simple way.

The second result is the identification of the relevant geometric
length scale of the target that we called the trapping length: $L =
|\Gamma|/C$.  This length naturally emerges from our approximation as
the geometric scale, to which the physical reaction length $1/q =
D/\kappa$ has to be compared with.  This trapping length generalizes a
former length $L_S = |\Gamma|/\diam\{\Gamma\}$, introduced by Sapoval
{\it et al.} \cite{Sapoval02}, to anisotropic targets and higher
dimensions.  The simple form of the trapping length is quite
intuitive.  In fact, the surface area $|\Gamma|$ naturally appears in
the reaction-limited regime ($q\to 0$) when the transport step is fast
as compared to the reaction step and thus the reaction event occurs on
any target point with almost equal probabilities (i.e., the so-called
spread harmonic measure is almost uniform, see
\cite{Grebenkov06c,Grebenkov15}).  For instance, the principal
eigenvalue exhibits the well-known behavior $\lambda_1^{(q)} \approx q
|\Gamma|/|\Omega|$.  In the opposite diffusion-limited regime
($q\to\infty$), the trapping capacity of the target is determined by
its capacity $C$, yielding $\lambda_1^{(q)} \approx C/|\Omega|$.  The
role of the capacity as the principal geometric characteristic of the
target can be recognized in the seminal paper by Smoluchowski
\cite{Smoluchowski1917}, in which the steady-state flux was shown to be
proportional to the radius of a spherical target, i.e., to its
capacity.  While the reaction length $1/q = D/\kappa$ is the ratio of
two transport coefficients, the trapping length $L = |\Gamma|/C$ is
the ratio of the associated geometric characteristics of the target.
In this light, our approximation (\ref{eq:lambda1}) can also be viewed
as an interpolation between two limiting regimes.  However, its
derivation and high accuracy suggest that Eq. (\ref{eq:lambda1})
correctly represents the dependence of the principal eigenvalue on the
main parameters of the problem, at least for small targets.

The third and last result concerns the target anisotropy, which was
mainly ignored in former studies.  We obtained the exact relations for
the trapping length of both prolate and oblate spheroids in $\R^d$
with $d \geq 3$ (an extension to more general bi-axial ellipsoids is
discussed in Appendix \ref{sec:biaxial}).  We showed that the trapping
length $L$ vanishes as an elongated target gets thinner.  As such a
target is hardly accessible to Brownian motion, one might expect to
deal with the diffusion-limited regime.  However, the vanishing of $L$
implies that diffusion-controlled reactions on needle-like targets are
always in the reaction-limited regime.  In other words, even though it
is hard to find such a target for the first time, it is even more
difficult to retrieve the target after each failed attempt to react.
In contrast, the trapping capacity of flattened (disk-like) targets is
not significantly different from round ones.

Our approximation is valid for any space dimension $d \geq 3$,
and its accuracy gets higher as $d$ grows.  It is therefore natural to
ask what happens in the planar case ($d = 2$), which stands apart by
several reasons.  In fact, the recurrent nature of Brownian motion in
the plane drastically changes many diffusive properties as compared to
higher-dimensional settings, for which Brownian motion is transient.
First, a steady-state solution of Eq. (\ref{eq:Psi_def}) that was
defined the harmonic capacity, does not exist for unbounded planar
domains.  This can be easily seen by considering a disk-shaped
capacitor $\C$, for which the problem (\ref{eq:Psi_def}) does not
depend on the angular coordinate.  A general radial solution of the
Laplace equation in polar coordinates, $\Delta u = \tfrac{1}{r}
\partial_r r \partial_r u = 0$, has a form $c_1 + c_2 \ln r$, and there is
no way to choose arbitrary constants $c_1$ and $c_2$ to get $u(r) \to
0$ as $r\to\infty$, except for the trivial solution with $c_1 = c_2 =
0$.  In particular, the probability of capture $\Psi(\x)$ is always
equal to $1$ for planar domains.  This particular issue can be
resolved by replacing the harmonic capacity by the logarithmic
capacity \cite{Garnett}.  The related asymptotic analysis was realized
in earlier works (see
\cite{Ozawa81,Mazya85,Ward93,Ward93b,Kolokolnikov05} and references
therein); in particular, an expansion of the principal eigenvalue in
powers of $\nu = 1/\ln(\ve)$ was derived, where $\ve$ is the relative
size of the target.  The major difference from higher-dimensional
settings is a very weak logarithmic dependence of the expansion
parameter $\nu$ on the relative target size $\ve$ so that the leading
order of the expansion is usually inaccurate, except for extremely
small targets.  In other words, one needs to deal with an expansion,
which contains many terms that are not easily accessible and depend on
various geometric properties of the confining domain and the the
target.  More generally, the logarithmic form of the fundamental
solution of the Laplace equation in the plane,
$-\ln(|\x-\x'|)/(2\pi)$, is responsible for ``long-range
interactions'' between distant points of space such as, for instance,
the strong impact of an outer boundary onto the behavior near the
target.  This fundamental difference makes our approach less useful in
the plane.

The present work has several perspectives and possible extensions.
First, it would be interesting to re-derive the approximation
(\ref{eq:lambda1}) in a more rigorous way and/or by a direct analysis
of the eigenvalue problem, e.g., by matched asymptotic methods.  In
fact, our derivation involved three approximations, and it was
difficult to control the accuracy and relevance of each step.  Second,
one can deal with multiple small targets.  If the sizes of targets are
much smaller than the distances between them and from the outer
reflecting boundary, the approximation (\ref{eq:lambda1}) is expected
to hold.  Note that the capacity of the union of small targets is
equal, in the leading order, to the sum of their capacities; the
surface area is also additive.  Moreover, Cheviakov and Ward derived
the next-order correction term to the principal eigenvalue for a
configuration of perfect targets \cite{Cheviakov11}.  This correction
term can be used to define the ``corrected'' capacity $C'$, as we did
in Eq. (\ref{eq:Cprime}) for a single target.  A numerical validation
of this approximation in configurations with multiple targets presents
an important perspective.  When the targets are spherical, one can
apply efficient semi-analytical methods based on addition theorems
(see \cite{Grebenkov19f,Grebenkov20b} and references therein).
Another validation step concerns irregularly-shaped targets, whose
surface area and thus the trapping length can be (arbitrarily) large,
despite their smallness.  Such a situation is not possible for
spheroids, for which $L \leq b/(d-2)$, see Fig. \ref{fig:ell}, i.e.,
the smallness of the target diameter $2b$ implied the smallness of
$L$.  The accuracy of our approximation for $L/b \gg 1$ remains to be
analyzed.  Finally, one can investigate other surface reaction
mechanisms (beyond the conventional Robin boundary condition) by using
an encounter-based approach
\cite{Grebenkov20a,Grebenkov20d,Grebenkov20g,Grebenkov22}.  Here, the
explicit dependence of the reactivity parameter $q$ may allow to
access various properties of diffusion-mediated surface phenomena.

\begin{acknowledgments}
D.S.G. acknowledges the Alexander von Humboldt Foundation for support
within a Bessel Prize award.
\end{acknowledgments}

\appendix
\section{Prolate spheroids}
\label{sec:prolate}

The harmonic capacity and the surface area of general ellipsoids in
$\R^d$ with $d > 3$ have been studied in \cite{Tee05}.  Here we
describe the main derivation steps and further simplifications that we
managed to get for prolate spheroids defined by
Eq. (\ref{eq:Gamma_prolate}), with $d-1$ minor semi-axes $a$ and one
major semi-axis $b$ such as $a < b$.  Combining the standard prolate
spheroidal coordinates in $\R^3$ with multidimensional spherical
coordinates, one can introduce the following $d$-dimensional
spheroidal coordinates:
\begin{align*}
x_{d} &= c \cosh(\alpha) \cos(\theta_1), \\
x_{d-1} &= c \sinh(\alpha) \sin(\theta_1) \cos(\theta_2), \\
x_{d-2} &= c \sinh(\alpha) \sin(\theta_1) \sin(\theta_2) \cos(\theta_3), \\
& \vdots \\
x_{2} &= c \sinh(\alpha) \sin(\theta_1) \ldots \sin(\theta_{d-2}) \cos(\phi), \\
x_1 &= c \sinh(\alpha) \sin(\theta_1) \ldots \sin(\theta_{d-2}) \sin(\phi),
\end{align*}
where $c = \sqrt{b^2-a^2}$ is the focal half-distance, $0 < \alpha <
\infty$ is analogous to the radial coordinate, whereas $0 \leq
\theta_i \leq \pi$ and $0 \leq \phi < 2\pi$ are angular coordinates.
Substituting these coordinates in the quadratic equation in
Eq. (\ref{eq:Gamma_prolate}), one set $\cosh(\alpha_0) = b/c$ (and
thus $\sinh(\alpha_0) = a/c$) to determine the ``radial'' coordinate
$\alpha_0$ of the spheroidal boundary $\Gamma_{a,b}$.  The following
construction is fairly standard in differential geometry
\cite{Dubrovin,Berger}.  In fact, one first determines the basis
vectors associated to new coordinates, e.g., the vector $\vec e_\alpha
= (dx_1/d\alpha,\ldots,dx_d/d\alpha)^{\dagger}$ is associated to
$\alpha$, etc.  The norms of these vectors determine the scale
factors:
\begin{align*}
h_\alpha & = h_{\theta_1} = c \sqrt{\sinh^2\alpha + \sin^2\theta_1} \, , \\
h_{\theta_k} & = c \, \sinh\alpha \, \sin \theta_1 \ldots \sin \theta_{k-1}  \quad (k=2,3,\ldots,d-2),  \\
h_{\phi} & = c \, \sinh\alpha \, \sin\theta_1 \ldots \sin \theta_{d-2} ,
\end{align*}
from which the metric, volume and surface elements, and the form of
the Laplace operator follow.  Skipping these technical details, we
write the Laplace operator as
\begin{align}  \nonumber
\Delta & = \frac{1}{c^2(\sinh^2\alpha + \sin^2\theta_1)} \left( \partial^2_\alpha + (d-2) \coth\alpha \partial_\alpha \right) \\  \nonumber
&	+\frac{1}{c^2(\sinh^2\alpha + \sin^2\theta_1)} \left( \partial^2_{\theta_1} + (d-2) \cot\theta_1 \partial_{\theta_1} \right) \\  \nonumber
&	+ \frac{1}{c^2 \sinh^2\alpha  \sin^2\theta_1} \left( \partial^2_{\theta_2} + (d-3) \cot\theta_2 \partial_{\theta_2} \right) \\  \nonumber
&	+ \frac{1}{c^2 \sinh^2\alpha  \sin^2\theta_1 \sin^2\theta_{2}} 
 	\left( \partial^2_{\theta_3} + (d-4) \cot\theta_3 \partial_{\theta_3} \right) \\   \nonumber
& + \ldots \\ \nonumber
&	+\frac{1}{c^2 \sinh^2\alpha  \sin^2\theta_1 ... \sin^2\theta_{d-3}} 
	\left( \partial^2_{\theta_{d-2}} + \cot\theta_{d-2} \partial_{\theta_{d-2}} \right) \\  
&	+ \frac{1}{c^2 \sinh^2\alpha  \sin^2\theta_1 ... \sin^2\theta_{d-2}} \partial^2_{\phi} .
\end{align}

In order to compute the capacity, one needs to solve the Dirichlet
boundary value problem:
\begin{equation}  \label{eq:Psi_def1}
\Delta \Psi(\x) = 0 \quad (\x\in\R^d \backslash \C), \qquad 
\left\{ \begin{array}{l} \Psi|_{\partial \C} = 1 , \\ 
\lim\limits_{|\x|\to\infty} \Psi(\x) = 0, \\ \end{array} \right.
\end{equation}
where $\C$ is the interior of the prolate spheroid surrounded by
$\Gamma_{a,b}$.  Since the boundary condition is constant, the
solution of this problem is invariant under rotations around the
coordinate axis $x_d$.  In spheroidal coordinates, the function
$\Psi(\x)$ thus depends only on the ``radial'' coordinate $\alpha$ so
that only the first term in the above Laplace operator remains
\begin{equation}
\frac{1}{c^2(\sinh^2\alpha + \sin^2\theta_1)} \left( \partial^2_{\alpha} + (d-2) \coth\alpha\, \partial_{\alpha} \right) \Psi(\alpha) = 0.
\end{equation}
Setting $\xi = \cosh\alpha$, this equation is reduced to
\begin{equation}  \label{eq:DeltaP}
(\xi^2-1) \partial^2_{\xi} \Psi + (d-1) \xi \partial_{\xi} \Psi = 0,
\end{equation}
subject to the Dirichlet boundary condition $\Psi(\xi_0) = 1$ with
$\xi_0 = \cosh(\alpha_0) = b/c$ and the regularity condition
$\Psi(\xi) \to 0$ as $\xi\to \infty$.  Setting $u(\xi) =
\partial_{\xi} \Psi(\xi)$, one integrates Eq. (\ref{eq:DeltaP})
to get $u(\xi) = c_1 (\xi^2 - 1)^{(1-d)/2}$, with an arbitrary
constant $c_1$.  The integral of this function yields
\begin{equation}  \label{eq:Psi_prolate0}
\Psi(\xi) = c_1 \int\limits_{\xi}^\infty dz \, \bigl(z^2 - 1\bigr)^{-\eta} \,, \qquad \eta = \frac{d-1}{2}\,,
\end{equation}
whose form ensures the regularity condition.  
Setting $y = 1/z^2$ and using the Taylor expansion of $(1-y)^{-\eta}$,
one can express this integral in terms of the hypergeometric function
\begin{align*}
\Psi(\xi) &  = \frac{c_1}{2} \int\limits_0^{1/\xi^2} dy \, y^{\eta-3/2}(1-y)^{-\eta} \\
& = c_1 \frac{\xi^{1-2\eta}}{2\eta-1} \, ~_2F_1(\eta,\eta-1/2; \eta+1/2; 1/\xi^2) \\
& = c_1 \frac{(\xi^2-1)^{1-\eta}}{\xi(2\eta-1)} \, ~_2F_1(1/2,1; \eta+1/2; 1/\xi^2).
\end{align*}
Substituting $\eta = (d-1)/2$ and $\xi_0 = \cosh\alpha_0 = b/c$, we
determine the constant $c_1$ from the Dirichlet boundary condition:
\begin{equation}
c_1 = \frac{(d-2)b}{a^{3-d} c^{d-2} \, ~_2F_1(1/2,1; d/2; c^2/b^2)} \,.
\end{equation}

Finally, we need to evaluate the integral of the normal derivative of
the solution in Eq. (\ref{eq:Psi_prolate0}),
\begin{equation}
(\partial_n \Psi)_{|\Gamma_{a,b}} = - \left(\frac{1}{h_\alpha} \partial_\alpha \Psi \right)_{\alpha = \alpha_0} \,,
\end{equation}
over the surface $\Gamma_{a,b}$:
\begin{align}  \nonumber
C^{(d)}_{a,b} &= \int\limits_{\Gamma_{a,b}} d\x \, (\partial_n \Psi) \\  \nonumber
& = c^{d-2} [\sinh \alpha_0]^{d-2} \int\limits_0^{\pi} d\theta_1 \, \sin^{d-2} \theta_1 
\int\limits_0^{\pi} d\theta_2 \, \sin^{d-3} \theta_2 \\
&  \ldots \int\limits_0^{\pi} d\theta_{d-2} \, \sin \theta_{d-2}
\int\limits_0^{2\pi} d\phi \, \bigl(-\partial_\alpha \Psi \bigr)_{\alpha = \alpha_0} ,
\end{align}
where the surface element was expressed in terms of the scale factors
and we used that the equal scale factors $h_\alpha$ and $h_{\theta_1}$
compensated each other.  The integrals over angular coordinates yield
the surface area $\sigma_d$ of the unit sphere in $\R^d$ so that
\begin{align}  \nonumber
C^{(d)}_{a,b} &= \sigma_d  c^{d-2} [\sinh \alpha_0]^{d-2} \bigl(-\partial_\alpha \Psi \bigr)_{\alpha = \alpha_0}  
 = \sigma_d  \, c^{d-2} \, c_1  \\   \label{eq:Cd_prolate_A}
& = \frac{(d-2) \sigma_d \, a^{d-3}\, b}{~_2F_1(1/2,1; d/2; c^2/b^2)} \,,
\end{align}
i.e., we arrive at Eq. (\ref{eq:Cd_prolate}).  To our knowledge, such
a compact expression for the capacity of the prolate spheroid in
$\R^d$ was not earlier reported.

The surface area of ellipsoids was derived in \cite{Tee05}.  In our
particular case, the general expression can be written as 
\begin{equation}
|\Gamma_{a,b}^{(d)}| = \frac{4\pi^{(d-1)/2} a^{d-3} b^2}{\Gamma(\frac{d-1}{2})} 
\int\limits_0^1 dx \, \frac{(1-x^2)^{(d-3)/2}}{(1+\delta x^2)^{(d+1)/2}} \,,
\end{equation}
where $\delta = b^2/a^2 - 1$.  This integral can be expressed in terms
of the hypergeometric function:
\begin{equation}  \label{eq:Gamma_prolate_auxil1}
|\Gamma_{a,b}^{(d)}| = \frac{2\pi^{d/2} a^{d-3} b^2}{\Gamma(d/2)} ~ _2F_1\biggl(\frac12,\frac{d+1}{2}; \frac{d}{2}; 1 - \frac{b^2}{a^2}\biggr).
\end{equation}
Using the Pfaff transformation, one can rewrite it as
Eq. (\ref{eq:area_prolate}).

In three dimensions, one retrieves the classical expression
\begin{equation}
|\Gamma_{a,b}^{(3)}| = 2\pi a^2 \biggl(1 + \frac{b}{ae} \sin^{-1} (e) \biggr),  \quad e = \sqrt{1 - a^2/b^2} ,
\end{equation}
so that the trapping length reads
\begin{equation}
L = \frac{a^2 \bigl(1 + \frac{b}{ae} \sin^{-1} (e) \bigr) \ln \bigl(\frac{1 + e}{1 - e}\bigr)}{4eb} \,.
\end{equation}
Note that $L \approx a \tfrac{\pi}{4} \ln(2b/a)$ as $a\to 0$.

\section{Oblate spheroids}
\label{sec:oblate}

The derivation for oblate spheroids is very similar.  One introduces
an extension of the oblate spheroidal coordinates as
\begin{align*}
x_{d} &= c \sinh(\alpha) \sin(\theta_1), \\
x_{d-1} &= c \cosh(\alpha) \cos(\theta_1) \sin(\theta_2), \\
x_{d-2} &= c \cosh(\alpha) \cos(\theta_1) \cos(\theta_2) \sin(\theta_3), \\
& \vdots \\
x_{2} &= c \cosh(\alpha) \cos(\theta_1) \ldots \cos(\theta_{d-2}) \sin(\phi), \\
x_1 &= c \cosh(\alpha) \cos(\theta_1) \ldots \cos(\theta_{d-2}) \cos(\phi),
\end{align*}
with $c = \sqrt{b^2-a^2}$, $0< \alpha < \infty$, $-\pi/2 \leq
\theta_i\leq \pi/2$, and $0 \leq \phi < 2\pi$.  These coordinates
determine the scale factors
\begin{align*}
h_\alpha & = h_{\theta_1} = c \sqrt{\sinh^2\alpha + \sin^2\theta_1}  , \\
h_{\theta_k} & = c \, \cosh\alpha \, \cos \theta_1 \ldots \cos \theta_{k-1}  \quad (k=2,3,\ldots,d-2),  \\
h_{\phi} & = c \, \cosh\alpha \, \cos\theta_1 \ldots \cos \theta_{d-2} ,
\end{align*}
from which the metric and the Laplace operator follow.  In particular,
the solution of the boundary value problem (\ref{eq:Psi_def1}) depends
only on the ``radial coordinate'' $\alpha$:
\begin{equation}
\frac{1}{c^2(\sinh^2\alpha + \sin^2\theta_1)} \left( \partial^2_{\alpha} + (d-2) \tanh\alpha\, \partial_{\alpha} \right) \Psi(\alpha) = 0.
\end{equation}
Setting $\xi = \sinh\alpha$, this equation is reduced to
\begin{equation}
(\xi^2+1) \partial^2_{\xi} \Psi + (d-1) \xi \partial_{\xi} \Psi = 0,
\end{equation}
subject to the Dirichlet boundary condition $\Psi(\xi_0) = 1$ with
$\xi_0 = \sinh(\alpha_0) = a/c$ and the regularity condition $\Psi(\xi)
\to 0$ as $\xi\to \infty$.  Setting $u(\xi) = \partial_{\xi} \Psi(\xi)$, one
gets $u(\xi) = c_1 (\xi^2 + 1)^{(1-d)/2}$, with an arbitrary constant
$c_1$.  The integral of this function yields
\begin{equation}  \label{eq:Psi_oblate0}
\Psi(\xi) = c_1 \int\limits_{\xi}^\infty dz \, \bigl(z^2 + 1\bigr)^{-\eta} \,, \qquad \eta = \frac{d-1}{2}\,.
\end{equation}
As previously, one can express this solution as
\begin{align}  \nonumber
\Psi(\xi) & = \frac{c_1}{2} \int\limits_0^{1/\xi^2} dy \, y^{\eta-3/2}(1+y)^{-\eta} \\  \nonumber
& = c_1 \frac{\xi^{1-2\eta}}{2\eta-1} \, ~_2F_1(\eta,\eta-1/2; \eta+1/2; -1/\xi^2) \\
& = c_1 \frac{(\xi^2+1)^{1-\eta}}{\xi(2\eta-1)} \, ~_2F_1(1/2,1; \eta+1/2; -1/\xi^2).
\end{align}
Substituting $\eta = (d-1)/2$ and $\xi_0 = \sinh(\alpha_0) = a/c$, we
get
\begin{equation}
c_1 = \frac{(d-2)a}{b^{3-d} c^{d-2} \, ~_2F_1(1/2,1; d/2; -c^2/a^2)} \,.
\end{equation}

To complete the computation, we need to evaluate the integral of the
normal derivative of this solution,
\begin{equation}
(\partial_n \Psi)_{|\tilde{\Gamma}_{a,b}} = - \left(\frac{1}{h_\alpha} \partial_\alpha \Psi \right)_{\alpha = \alpha_0} \,,
\end{equation}
over the surface $\tilde{\Gamma}_{a,b}$:
\begin{align}  \nonumber
\tilde{C}^{(d)}_{a,b} &= \int\limits_{\tilde{\Gamma}_{a,b}} d\x \, (\partial_n \Psi) \\  \nonumber
& = c^{d-2} [\cosh \alpha_0]^{d-2} \hspace*{-2mm} \int\limits_{-\pi/2}^{\pi/2} \hspace*{-1mm} d\theta_1 \, \cos^{d-2} \theta_1 
\hspace*{-2mm}  \int\limits_{-\pi/2}^{\pi/2} \hspace*{-1mm} d\theta_2 \, \cos^{d-3} \theta_2 \\
&  \ldots \int\limits_{-\pi/2}^{\pi/2} d\theta_{d-2} \, \cos \theta_{d-2}
\int\limits_0^{2\pi} d\phi \, \bigl(-\partial_\alpha \Psi \bigr)_{\alpha = \alpha_0} .
\end{align}
Evaluating the integrals over angular coordinates, we get
\begin{align}  \nonumber
\tilde{C}^{(d)}_{a,b} &= \sigma_d  c^{d-2} [\cosh \alpha_0]^{d-2} \bigl(-\partial_\alpha \Psi \bigr)_{\alpha = \alpha_0}  
 = \sigma_d  \, c^{d-2} \, c_1  \\   
& = \frac{(d-2) \sigma_d \, a \, b^{d-3}}{~_2F_1(1/2,1; d/2; -c^2/a^2)} \,.
\end{align}
Using the Pfaff transformation, one can rewrite this expression as
Eq. (\ref{eq:Coblate}).  To our knowledge, such a compact expression
for the capacity of the oblate spheroid in $\R^d$ was not earlier
reported.

The surface area of oblate spheroids is given by the formula
(\ref{eq:Gamma_prolate_auxil1}), in which $a$ and $b$ are exchanged:
\begin{equation}  \label{eq:Gamma_oblate_auxil1}
|\tilde{\Gamma}_{a,b}^{(d)}| = \sigma_d \, b^{d-3} a^2 ~ _2F_1\biggl(\frac12,\frac{d+1}{2}; \frac{d}{2}; 1 - \frac{a^2}{b^2}\biggr).
\end{equation}
Using the Euler transformation, one gets a more convenient
representation (\ref{eq:area_oblate}).

In three dimensions, one retrieves the classical formula
\begin{equation}
|\tilde{\Gamma}_{a,b}^{(3)}| = 2\pi b^2  + \pi \frac{a^2}{e} \ln \frac{1+e}{1-e} \, .
\end{equation}
The trapping length is
\begin{equation}
L = \frac{2\pi b^2  + \pi \frac{a^2}{e} \ln \frac{1+e}{1-e}}{4\pi c} \, \cos^{-1}(a/b) \, .
\end{equation}

\section{Bi-axial ellipsoids}
\label{sec:biaxial}

The prolate and oblate spheroids discussed in Appendices
\ref{sec:prolate} and \ref{sec:oblate} are particular cases of a
bi-axial ellipsoid, which has $p$ minor semi-axes $a$ and $q$ major
semi-axes $b$ (such that $a < b$).  For the sake of completeness, we
provide here the exact expressions for the capacity and the surface
area of these domains.  We recast former results by Tee in
\cite{Tee05} in a simpler form in terms of hypergeometric functions.

Tee obtained the following formula for the capacity of a bi-axial
ellipsoid with $p$ minor semi-axes $a$ and $q$ major semi-axes $b >
a$:
\begin{equation}
\frac{1}{C} = \frac{1}{b^{p+q-2} \sigma_{p+q}} \int\limits_0^1 dx \frac{x^{p+q-3}}{\bigl(1-(1-a^2/b^2) x^2\bigr)^{p/2}} \,,
\end{equation}
where $\sigma_{d}$ is given by Eq. (\ref{eq:omegad}).  Expanding the
denominator into a Taylor series of powers of $x$, we get
\begin{equation}  \label{eq:Cbiaxial1}
C = \frac{(p+q-2) \sigma_{p+q} b^{p+q-2}}{_2F_1\bigl(\frac{p}{2}, \frac{p+q-2}{2}; \frac{p+q}{2}; 1-a^2/b^2\bigr)} \,.
\end{equation}
The Euler transformation allows one to get another representation:
\begin{equation}  \label{eq:Cbiaxial2}
C = \frac{(p+q-2) \sigma_{p+q}  a^{p-2} b^{q}}{_2F_1\bigl(\frac{q}{2}, 1; \frac{p+q}{2}; 1-a^2/b^2\bigr)} \,.
\end{equation}
For instance, setting $p = d-1$ and $q = 1$ into the last formula, we
retrieve Eq. (\ref{eq:Cd_prolate_A}) for a prolate spheroid in $\R^d$.
Similarly, setting $p = 1$ and $q = d-1$ into Eq. (\ref{eq:Cbiaxial1})
yields Eq. (\ref{eq:Coblate}) for an oblate spheroid.

Tee expressed the surface area of bi-axial ellipsoids in terms of the
integrals
\begin{equation}
I_{\alpha,\beta}(\delta) = \int\limits_0^1 dh \, \frac{(1-h^2)^\alpha}{(1-\delta h^2)^\beta} \,.
\end{equation}
Setting $\mu = \delta/(\delta - 1)$ and using the Taylor expansion of
$(1-\mu x)^{-\beta}$, we have
\begin{align}  \nonumber
& I_{\alpha,\beta}(\delta) = \frac{1}{2(1-\delta)^\beta} \int\limits_0^1 \frac{dx}{\sqrt{1-x}} \, \frac{x^\alpha}{(1- \mu x)^\beta} \\  \label{eq:Ialpha}
& = \frac{\sqrt{\pi}\, \Gamma(\alpha+1)}{2(1-\delta)^\beta \Gamma(\alpha+\frac{3}{2})} ~
 _2F_1\biggl(\beta, \alpha+1;\alpha+\frac{3}{2}; \frac{\delta}{\delta-1}\biggr),
\end{align}
where we used
\begin{equation*}
\int\limits_0^1 dx \frac{x^\alpha}{\sqrt{1-x}} = \frac{\sqrt{\pi} \, \Gamma(\alpha+1)}{\Gamma(\alpha+3/2)} \,.
\end{equation*}

Depending on the parity of $p$ and $q$, Tee treated separately three
cases and expressed the surface area of the corresponding bi-axial
ellipsoids in terms of $I_{\alpha,\beta}(\delta)$, with $\alpha$ and
$\beta$ being related to $p$ and $q$.  Using Eq. (\ref{eq:Ialpha}), we
managed to show that all three cases yield the same result.  Skipping
technical details of this analysis, we provide the following exact
expression for the surface area:
\begin{align}  \nonumber
|\Gamma| & = \frac{\sigma_{p+q}  a^{p-1} b^{q}}{p+q-1} \biggl\{ \frac{(q-1)a^2}{b^2} ~ _2F_1\biggl(\frac{1}{2}, \frac{q}{2}; 
\frac{p+q}{2}; 1- \frac{a^2}{b^2}\biggr) \\  \label{eq:Gamma_biaxial}
& + p ~ _2F_1\biggl(\frac{1}{2}, \frac{q}{2}-1; \frac{p+q}{2}; 1- \frac{a^2}{b^2}\biggr)\biggr\} .
\end{align}
For a prolate spheroid with $q = 1$ and $p = d-1$, we retrieve
Eq. (\ref{eq:area_prolate}).  For an oblate spheroid with $q = d-1$
and $p = 1$, one can use contiguous relations between hypergeometric
functions to retrieve Eq. (\ref{eq:area_oblate}).

Substituting Eq. (\ref{eq:Cbiaxial1}) or Eq. (\ref{eq:Cbiaxial2}) for
$C$ and Eq. (\ref{eq:Gamma_biaxial}) for $|\Gamma|$ into
Eq. (\ref{eq:ell}), one determines the trapping length of a general
bi-axial ellipsoid.

\section{Numerical solution by finite elements method}
\label{sec:FEM}

\begin{figure}[h!]
\begin{center}
\includegraphics[width=40mm]{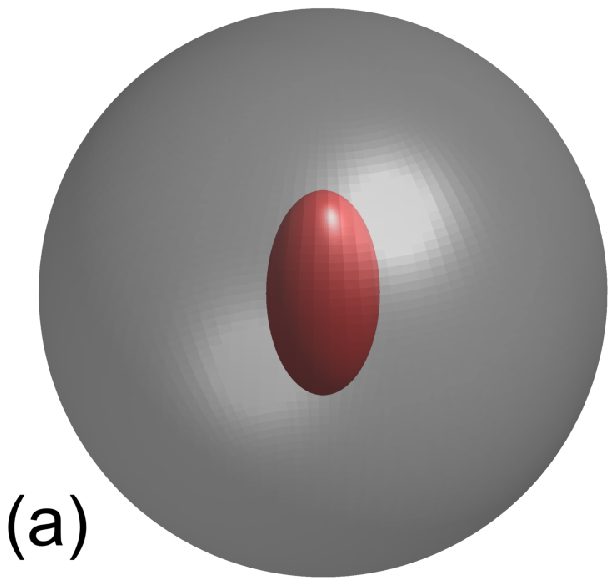} 
\includegraphics[width=44mm]{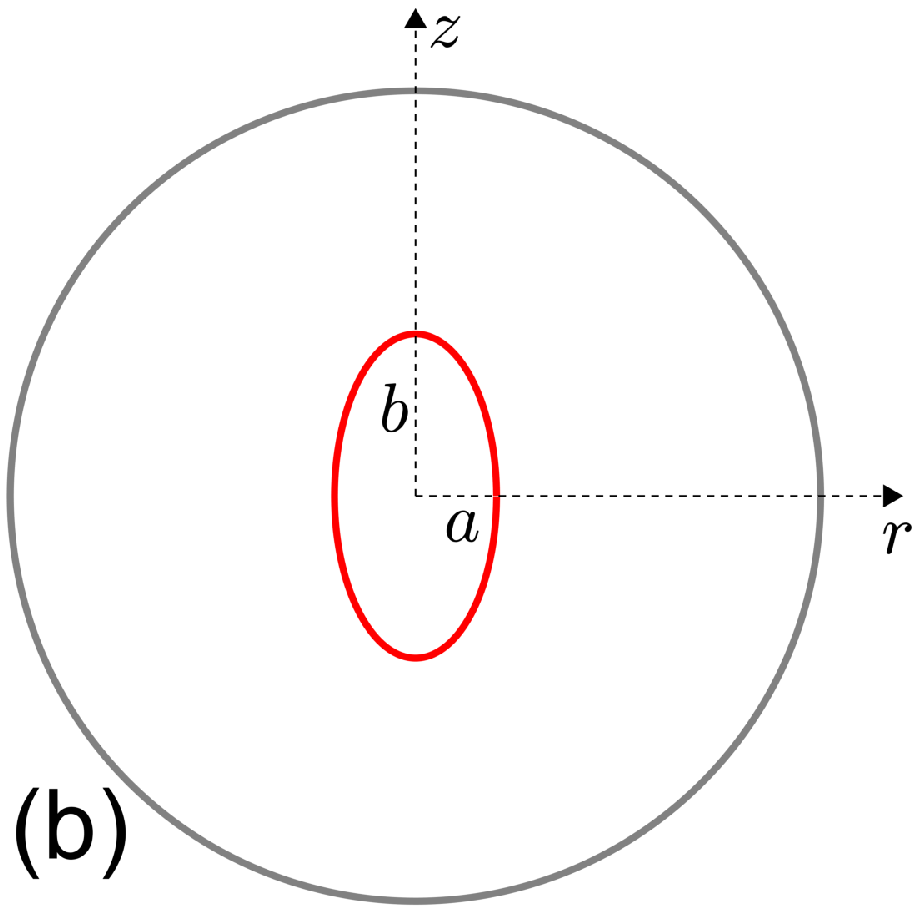} 
\end{center}
\caption{
{\bf (a)} A prolate spheroidal target (in red) is enclosed by an outer
reflecting sphere (in gray).  {\bf (b)} An equivalent planar domain
with an elliptic target (in red) and an outer circular reflecting
boundary (in gray). }
\label{fig:scheme_FEM}
\end{figure}

In order to check the accuracy of our approximation, we solved the
underlying boundary value problem by a finite elements method.  The
axial symmetry of spheroids allowed us to reduce the original
$d$-dimensional problem to a planar one.  In fact, one can write the
Laplace operator in the cylindrical coordinates as
\begin{equation}
\Delta = \partial_z^2 + \frac{1}{r^{d-2}} \partial_r \, r^{d-2} \, \partial_r  + \frac{1}{r^2} \Delta_{\rm ang} \,,
\end{equation}
where $z$ denotes the coordinate along the symmetry axis (i.e., $z =
x_d$), $r = \sqrt{x_1^2 + \ldots + x_{d-1}^2}$, and $\Delta_{\rm ang}$
is the angular part of the Laplace operator in the hyperplane
$\R^{d-1}$, which is orthogonal to the axis $x_d$.  As the original
eigenvalue problem in Eq. (\ref{eq:un}) is invariant under rotations
along the $x_d$ axis, its solution does not depend on the angular
part.  It can thus be written as
\begin{equation}
- \nabla c \nabla u = \lambda r^{d-2} u ,
\end{equation}
where $\nabla$ is the gradient operator in the $(r,z)$ plane, and $c$
is the diagonal $2\times 2$ matrix with entries $r^{d-2}$.  This
reduced eigenvalue problem has to be solved in the planar
cross-section of the domain (see Fig. \ref{fig:scheme_FEM}).  The
problem was solved numerically by PDETool in Matlab.  We compared
numerical solutions with different choices for the maximal meshsize to
ensure that the results do not depend on this choice.

\end{document}